\documentclass[prd,floatfix,nofootinbib,preprintnumbers]{revtex4}
\usepackage{epsfig}
\usepackage{graphicx}
\usepackage{amsfonts}
\usepackage{mathrsfs}
\usepackage{psfrag}
\newcommand{\be}{\begin{equation}}
\newcommand{\ee}{\end{equation}}
\newcommand{\bea}{\begin{eqnarray}}
\newcommand{\eea}{\end{eqnarray}}
\newcommand{\nnb}{\nonumber}

\def\lsim{\raise0.3ex\hbox{$\;<$\kern-0.75em\raise-1.1ex\hbox{$\sim\;$}}}
\def\gsim{\raise0.3ex\hbox{$\;>$\kern-0.75em\raise-1.1ex\hbox{$\sim\;$}}}

\begin{document}
\title{Exclusive Semileptonic Rare Decays $B \to K^{(*)} l^+ l^-$ in a SUSY SO(10) GUT}
\author{Wen-Jun Li$^{a,b}$, Yuan-Ben Dai$^{a}$, Chao-Shang Huang$^{a}$ }
\affiliation{ $^{a}$Institute of Theoretical Physics, Academia
Sinica, P. O. Box 2735, Beijing 100080, China\\ $^{b}$Graduate
School of the Chinese Academy of Science, YuQuan Road 19A, Beijing
100039, China }

\begin{abstract}

   In the SUSY SO(10) GUT context, we study the exclusive processes
$B \to K^{(*)} l^+l^-(l=\mu,\tau)$. Using the Wilson coefficients
of relevant operators including the new operators
$Q_{1,2}^{(\prime)}$ which are induced by neutral Higgs boson
(NHB) penguins, we evaluate some possible observables associated
with these processes like, the invariant mass spectrum (IMS),
lepton pair forward backward asymmetry (FBA), lepton polarization
asymmetries etc. In the model the contributions from Wilson
coefficients $C_{Q_{1,2}}^\prime$, among new contributions, are
dominant. Our results show that the NHB effects are sensitive to
the FBA, $dL/d\hat{s}$, and $dT/d\hat{s}$ of $B \to K^{(*)} \tau^+
\tau^-$ decay, which are expected to be measured in B factories,
the deviation of $dT/d\hat{s}$ in $B \to K \mu^+\mu^-$ can reach
0.1 from SM, which could be seen in B factories, and the average
of the normal polarization $dN/d\hat{s}$ can reach several percent
for $B \to K \mu^+ \mu^-$ and it is 0.05 or so for $B\to K
\tau^+\tau^-$, which could be measured in the future super B
factories and provide a useful information to probe new physics
and discriminate different models.
\end{abstract}

\maketitle \noindent
\section{Introduction}
      The rich flavor changing neutral current processes  $B \to K^{(*)}l^+l^-$ have
been the sharper focus since these decays are  potential testing
grounds for the SM at loop level and hoped to probe the new
physics beyond the SM. Recently exclusive measurements have been
done by Belle and BaBar and the following results for the
branching ratios of the $B \to K \ell^+ \ell^-$ and $B \to K^\ast
\ell^+ \ell^-$ ($l=e, \mu$) decays are
announced~\cite{ichep,Aubert}: \bea {\cal B}r(B \to K \ell^+
\ell^-) = \left\{
\begin{array}{lc} \left( 4.8^{+1.0}_{-0.9} \pm 0.3 \pm 0.1\right)
\times
10^{-7}& [Belle]~,\\ \\
\left( 0.65^{+0.14}_{-0.13} \pm 0.04 \right) \times 10^{-6}&
[BaBar]~,\end{array} \right.  \eea \bea {\cal B}r(B \to K^\ast
\ell^+ \ell^-) = \left\{
\begin{array}{lc} \left( 11.5^{+2.6}_{-2.4} \pm 0.8 \pm 0.2\right)
\times 10^{-7}& [Belle]~,\\ \\ \left( 0.88^{+0.33}_{-0.29} \right)
\times 10^{-6}& [BaBar]~,\end{array} \right. \nnb \eea  which
imply
\begin{eqnarray}\label{exp}
Br(B \to  K l^+l^-)_{\rm
world-ave}&=& (0.54\pm 0.09) \times 10^{-6} \nnb\\
Br(B \to  K^* l^+l^-)_{\rm world-ave}&=& (1.04\pm 0.22)\times
10^{-6}
\end{eqnarray}
In addition, the FBA for $B \to K^* l^+l^-$ have firstly observed
at Belle collider~\cite{ichep}.

The $B \to K^{(*)} l^+ l^-$ decays, induced by $b \to s l^+l^-$
transition at quark level, are experimentally easier to measure
than inclusive processes $B \to X_s l^+ l^-$. From the theoretical
point of view there are large uncertainties, which come mostly
from the decay form factors, to make predictions for the exclusive
processes. At present, the knowledge of form factors lacks a
precisely non-perturbative solution. A number of papers are
dedicated to calculating the form factors with various appropriate
methods~\cite{quarkmodel,svzqcd,lcsr,rcqm}. Among them, the QCD
sum rules on the light-cone (LCSRs), which deals with form factors
at small values of $\hat{s}$, the momentum transfer to leptons, is
a complementary to lattice approach and has the consistence with
perturbative QCD and the heavy quark limit. In this paper, we will
use the form factors calculated by the LCSRs~\cite{9910221}.

   The measurement of invariant mass spectrum, forward-backward
asymmetry, and lepton polarizations are efficient tools to
establish the new physics. There are a great deal of studies for
processes $B \to K^{(*)} l^+ l^-$ in theory. A model independent
analysis has been carried out in~\cite{modindep,dan} and a lot of
papers perform the investigation in many new physics
scenarios~\cite{9705222,9910221,0004262,0008210,0009149-2,
0003188,0112149,0112300,0204219cpp,0307276FBA} and some
works~\cite{0304084,0209228,double} are dedicated to the double
lepton polarization. It has been pointed
out~\cite{9705222,0004262,0009149-2} that in the some types of
two-Higgs-double model and SUSY models the neutral Higgs bosons
have sizable contributions to these decays (for $l=\mu, \tau$) at
large $\tan{\beta}$.
In~\cite{9910221}, Ali {\it et.al} calculate these quantities in
five scenarios of supersymmetric model with assuming no additional
phase. Kruger's studies~\cite{0008210} are focus on the $CP $
violation of $\bar{B} \to K l^+ l^-$ in the model with additional
$CP$ phases and an extended operator basis. In ref.~\cite{0004262}
only the Higgs penguins with chargino-stop propagated in the loop
are considered.

The rapid progress in neutrino experiment~\cite{neu} requires new
physics to provide a theoretical explanation. Motivated by
neutrino observations, a number of SUSY SO(10) models have been
proposed~\cite{asy,cmc1,0205111,bsv} and some phenomenological
consequences of the models have been
discussed~\cite{asy,bi,cmc2,0312145,0407263}. In SUSY SO(10) GUT
models, there is a complex flavor non-diagonal down-type squark
mass matrix element of 2nd and 3rd generations of order one in the
RR sector (i.e., $\delta_{23}^{dRR}$ non-zero with
$\delta_{23}^{dRR}\equiv (M^2_{\tilde{d}RR})_{23} /
m_{\tilde{q}}^2$, where $(M^2_{\tilde{d}RR})_{ij}$ is the flavor
non-diagonal squared right-handed down squark mass matrix element
and $m_{\tilde{q}}$ is the average right-handed down-type squark
mass) at the GUT scale~\cite{0205111} which can induce large
flavor off-diagonal couplings such as the coupling of gluino to
the quark and squark which belong to different generations. These
couplings are in general complex and may contribute to the process
of flavor changing neutral currents(FCNC). For specific, we use
the SUSY SO(10) model described in ref.~\cite{0205111}. The
details and a simple description of this model can be found in
Refs.~\cite{0205111,0407263}. In this paper, we investigate
exclusive decay $B \to K^{(*)} l^+ l^-(l=\mu,\tau)$ in the context
of SUSY SO(10) GUT. It is well-known that the effects of the
counterparts of usual chromo-magnetic and electro-magnetic dipole
moment operators as well as semileptonic operators with opposite
chirality are suppressed by ${m_s}/{m_b}$ and consequently
negligible in SM. However, in SUSY SO(10) GUTs their effects can
be significant, since $\delta_{23}^{dRR}$ can be as large as
0.5~\cite{0205111}. Furthermore, $\delta_{23}^{dRR}$ can induce
new operators, the counterparts of usual scalar operators
($Q_{1,2}$, for their definitions, see below) in SUSY models, due
to NHB penguins with gluino-down type squark propagated in the
loop. We include the contributions of these counterpart operators
and find that indeed they are dominant in the SUSY SO(10), using
the MIA with double insertions to calculate Wilson coefficients of
operators. The aim of our paper is make an analysis of the SUSY
contributions, in particular, the contributions of neutral Higgs
bosons, to the exclusive decay $B \to K^{(*)} l^+ l^-(l=\mu,\tau)$
in the context of SUSY SO(10) GUT.

    The paper is organized as follows. In section 2, we present
the effective Hamiltonian and hadronic matrix elements of relevant
operators in terms of form factors. In section 3, the expressions
of observables are given. In section 4, we give the sparticle mass
spectrum using the revised ISAJET. We make the numerical analysis
and draw the conclusion in section 5.

\section{Effective Hamiltonian and Form Factors}
   In the SUSY SO(10) GUT, after integrating the heavy degree of freedom from
   the full theory, the general effective Hamiltonian for $b \to s l^+ l^- $ can be written as follows:
\begin{eqnarray}\label{Heff}
 {\cal H}_{\rm eff} &=& -\frac{4G_F}{\sqrt2}
   V_{tb} V^*_{ts} \bigg[\!\sum_{i=1}^2 C_i(\mu)\,O_i(\mu)
   +\!\sum_{i=3}^{10}(C_i(\mu)O_i(\mu) + C_i^\prime(\mu)\,O_i^\prime(\mu))
   + \!\sum_{i=1}^{8}\!( C_i(\mu)\,Q_i(\mu)+ C_i^\prime(\mu)\,Q_i^\prime(\mu))\bigg]
   \nonumber \\&&  \,
\end{eqnarray}
  where $O_i(\mu)(i=1,\cdots,10)$ are dimension-six operators
and $C_i(\mu)$ are the corresponding Wilson coefficients at the
scale $\mu$~\cite{goto}. The additional operators
$Q_i(i=1,\cdots,8)$ come from the neutral Higgs exchange diagrams
and their definitions are given as~\cite{dyb,0009149-2}
\begin{eqnarray}\nonumber
Q_1&=&\frac{e^2}{16\pi^2}(\bar{s}^{\alpha}_Lb^{\alpha}_R)(\bar{l}
l)\\\nonumber
Q_2&=&\frac{e^2}{16\pi^2}(\bar{s}^{\alpha}_Lb^{\alpha}_R)(\bar{l}\gamma_5
l)\\\nonumber
Q_{3(4)}&=&\frac{g^2}{16\pi^2}(\bar{s}^{\alpha}_Lb^{\alpha}_R)(\sum_q\bar{q}^{\beta}
_{L(R)}q^{\beta}_{R(L)})\\\nonumber
Q_{5(6)}&=&\frac{g^2}{16\pi^2}(\bar{s}^{\alpha}_Lb^{\beta}_R)(\sum_q\bar{q}^{\beta}
_{L(R)}q^{\alpha}_{R(L)})\\\nonumber
Q_7&=&\frac{g^2}{16\pi^2}(\bar{s}^{\alpha}_L\sigma^{\mu\nu}b^{\alpha}_R)
(\sum_q\bar{q}^{\beta}_L\sigma_{\mu\nu}q^{\beta}_R)\\\nonumber
Q_8&=&\frac{g^2}{16\pi^2}(\bar{s}^{\alpha}_L\sigma^{\mu\nu}b^{\beta}_R)
(\sum_q\bar{q}^{\beta}_L\sigma_{\mu\nu}q^{\alpha}_R)\\
\label{eq:ehn}
\end{eqnarray}
and the corresponding Wilson coefficients can be found in
~\cite{wuxh} . The primed operators, the counterpart of the
unprimed operators, are obtained by replacing the chiralities in
the corresponding unprimed operators with opposite ones. The
explicit expressions of the operators governing $B \to K^{(*)}
l^+l^-$ are given as:
\begin{eqnarray}\label{operator}
O_7 &=& \frac{e}{16\pi^2}m_b(\bar{s}\sigma_{\mu \nu}P_Rb)F^{\mu
\nu}, \hspace{1cm} O'_7 = \frac{e}{16\pi^2}m_b(\bar{s}\sigma_{\mu
\nu}P_Lb)F^{\mu \nu}, \nonumber\\
O_9 &=&
\frac{e^2}{16\pi^2}(\bar{s}\gamma_{\mu}P_Lb)(\bar{l}\gamma^{\mu}l),
\hspace{1.2cm}
O'_9 = \frac{e^2}{16\pi^2}(\bar{s}\gamma_{\mu}P_Rb)(\bar{l}\gamma^{\mu}l), \nonumber\\
O_{10} &=&
\frac{e^2}{16\pi^2}(\bar{s}\gamma_{\mu}P_Lb)(\bar{l}\gamma^{\mu}\gamma_5l),
\hspace{1cm}
O'_{10} = \frac{e^2}{16\pi^2}(\bar{s}\gamma_{\mu}P_Rb)(\bar{l}\gamma^{\mu}\gamma_5l), \nonumber\\
Q_1 &=& \frac{e^2}{16\pi^2}(\bar{s}P_Rb)(\bar{l}l), \hspace{2cm}
Q'_1 =
\frac{e^2}{16\pi^2}(\bar{s}P_Lb)(\bar{l}l),\nonumber\\
Q_2 &=& \frac{e^2}{16\pi^2}(\bar{s}P_Rb)(\bar{l}\gamma_5 l),
\hspace{2cm} Q'_2 =
\frac{e^2}{16\pi^2}(\bar{s}P_Lb)(\bar{l}\gamma_5l)
\end{eqnarray}
where $P_{L,R}= (1\mp \gamma_5)/2$. From the above Hamiltonian, we
get the decay amplitude of $b \to s l^+ l^-$:
\begin{eqnarray}\label{amplitude}
   {\cal M}(b \to s l^+l^-)=&-&\frac{G_F\alpha}{\sqrt{2}\pi}V_{tb}V^*_{ts}\left\{C_9^{eff}
   [\overline{s}\gamma_{\mu}Lb][\overline{l}\gamma^{\mu}l]+
    C_{10}[\overline{s}\gamma_{\mu}Lb][\overline{l}\gamma^{\mu}\gamma_5l]\right.\nonumber \\
  &&\left.
  -2\hat{m}_bC_7^{eff}[\overline{s}i\sigma_{\mu \nu}\frac{\hat{q}^{\nu}}{\hat{s}}
   Rb][\overline{l}\gamma^{\mu}l]+C_{Q_1}[\overline{s}Rb][\overline{l}l]+
   C_{Q_2}[\overline{s}Rb][\overline{l}\gamma_5l]
   +(C_i(m_b)\leftrightarrow C'_i(m_b) )\right\}
\end{eqnarray}
where $s=q^2, \hat{s}=\frac{s}{m_B^2}$, $q=p_B-p_{K^{(*)}}$ is the
momentum transfer. The Wilson coefficient $C^{eff}_9(\mu)$ and
$C_7^{eff}$ are defined as:
\begin{eqnarray}
C_9^{eff}(\mu,\hat{s})&=&C_9(\mu)+Y(\mu,\hat{s})+\frac{3\pi}{\alpha^2}C(\mu)
\Sigma_{V_i=\psi(1s)...\psi(6s)}k_i\frac{\Gamma(V_i \to l^+
l^-)m_{V_i}}{m_{V_i}^2-\hat{s}m_B^2-im_{V_i}\Gamma_{V_i}},\label{c9}\\
C_7^{eff}&=&C_7-C_5/3-C_6\label{c7eff}
\end{eqnarray}
where $C^{eff}_9(\mu)$ contains the long distance effects
associated with real $\bar{c}c$ in the intermediate states $B \to
KJ/\psi(\psi') \to K l^+ l^-$, which can be expressed as the last
term in Eq.(\ref{c9}), as well as the short distance
contributions. The function $Y(\mu,\hat{s})$ comes from the
one-loop contributions of the four-quark operators and its
explicit expression can be found in~\cite{buras}. The
$C^{'eff}_9(\mu)$ and $C_7^{'eff}$ can be obtained by replacing
the unprimed Wilson coefficients with the corresponding primed
ones in the above formula.

      In virtue of the form factors in~\cite{9910221},
the hadronic matrix elements in the $B \to K l^+ l^-$ decay can be
expressed as:
\begin{eqnarray}\label{ffbkll}
 <K(p)|\overline{s}\gamma_{\mu}b|B(p_b)>&=&f_+(s)p_\mu
 +f_-(s)q_\mu,\\
<K(p)|\overline{s}\sigma_{\mu\nu}q^\nu(1+\gamma_5)b|B(p_b)>
&=&i\left\{p_\mu
s-q_\mu(m^2_B-m_K^2)\right\}\frac{f_T(s)}{m_B+m_K}
\end{eqnarray}
Using equations of motion, we obtain
\begin{eqnarray} <K(p_K)|\bar{s}b|B(p_B)>
 &=& \frac{m_B^2-m_K^2}{m_s-m_b}f_0(s)\;.
\end{eqnarray}
For $B \to K^* l^+ l^-$, the form factors are defined as follows.
\begin{eqnarray}\label{ffbksll}
<K^*(p_{K^*},\varepsilon)|\bar{s}\gamma_{\mu}(1\pm
\gamma_5)b|B(p_B)> &=&\epsilon_{\mu\nu\rho\sigma}\epsilon^{*\nu}
p^\rho_Bp_{K^*}^\sigma\frac{2V(s)}{m_B+m_{K^*}} \pm
i\epsilon^*_\mu(m_B+m_{K^*})A_1(s) \nonumber \\
&&\mp ip_\mu (\epsilon^*p_B)\frac{A_2(s)}{m_B+m_{K^*}} \nonumber \\
&&\mp iq_\mu(\epsilon^* p_B)\frac{2m_{K^*}}{s}(A_3(s)-A_0(s)), \\
<K^*(p_{K^*},\varepsilon)|\overline{s}\sigma_{\mu\nu}q^\nu(1 \pm
\gamma_5)b|B(p_B)>&=&i\epsilon_{\mu\nu\rho\sigma}\epsilon^{*\nu}
p^\rho_Bp_{K^*}^\sigma2T_1(s)\pm \epsilon^*_\mu
T_2(s)(m^2_B-m^2_{K^*})\nonumber \\
&&\mp(\epsilon^*p_B)p_\mu
\left(T_2(s)+T_3(s)\frac{s}{m^2_B-m^2_{K^*}}\right)
\pm(\epsilon^*p_B) q_\mu T_3(s)
\end{eqnarray}
and \begin{eqnarray}\label{ffbksll2}
<K^*(p_{K^*},\varepsilon)|\bar{s}(1\pm \gamma_5)b|B(p_B)> &=& \mp
i(\epsilon^* p_B)\frac{2m_{K^*}}{m_b+m_s} A_0(s) \end{eqnarray} by
means of equations of motion.

The $q^2$ dependence of the form factors can be parameterized as
$$F(\hat{s})= F(0)exp(c_1\hat{s}+ c_2\hat{s}^2+ c_3\hat{s}^3)$$
where related parameters are given in the Table.4
of~\cite{9910221}.
\section{The formula for observables}

   From Eq.(\ref{Heff}-\ref{ffbksll2}),
we can write the decay matrix elements asㄩ
\begin{eqnarray}\label{nhbs}
{\cal A}
=-\frac{G_F\alpha}{2\sqrt{2}\pi}V_{tb}V^*_{ts}m_B\left[T^1_\mu(\overline{l}\gamma^{\mu}l)
  +T^2_\mu (\overline{l}\gamma^{\mu}\gamma_5l)+S(\overline{l}l)
   \right\}
\end{eqnarray} where for $B \to K l^+ l^-$ decay:
 \begin{eqnarray}
 T^1_\mu&=&A'(\hat{s})\hat{p}_\mu,\nonumber \\
 T^2_\mu&=&C'(\hat{s})\hat{p}_\mu+D'(\hat{s})\hat{q}_\mu,\nonumber \\
 S&=&S_1(\hat{s})
\end{eqnarray}
 and for $B \to K^*l^+l^-$ decay:
 \begin{eqnarray}
 T^1_\mu &=&
 A(\hat{s})\epsilon_{\mu\rho\alpha\beta}\epsilon^{*\rho}
 \hat{p}^\alpha_B\hat{p}^\beta_{K^*}
 -iB(\hat{s})\epsilon^{*}_{\mu}+iC(\hat{s})(\epsilon^{*} \cdot
 \hat{p}_B)\hat{p}_{\mu},\nonumber \\
  T^2_\mu &=&E(\hat{s})\epsilon_{\mu\rho\alpha\beta}\epsilon^{*\rho}
  \hat{p}^\alpha_B\hat{p}^\beta_{K^*}-iF(\hat{s})\epsilon^*_\mu
 +iG(\hat{s})(\epsilon^* \cdot \hat{p}_B)\hat{p}_\mu
  +iH(\hat{s})(\epsilon^{*} \cdot \hat{p}_B)\hat{q}_\mu, \nonumber \\
 S&=&i2\hat{m}_{K^*}(\epsilon^* \cdot \hat{p}_B)S_2(\hat{s})
\end{eqnarray}
  where $p=p_B+p_{K^{(*)}},q=p_B-p_{K^{(*)}},\hat{m}=\frac{m}{m_B}$,
$\hat{p}=\frac{p}{m_B}$, and the auxiliary functions are defined
as:
\begin{eqnarray}\label{aux}
A'(\hat{s})&=&[C_9^{eff}(\hat{s})
+C_9^{'eff}(\hat{s})]f_+(\hat{s})
+\frac{2\hat{m}_b}{1+\hat{m}_K}(C^{eff}_7+C^{'eff}_7)f_T(\hat{s}),\\
C'(\hat{s})&=&(C_{10}+C'_{10})f_+(\hat{s}),\\
D'(\hat{s})&=&(C_{10}+C'_{10})f_-(\hat{s})-
\frac{1-\hat{m}^2_K}{2\hat{m}_l(\hat{m}_b-\hat{m}_s)}(C_{Q_2}+C'_{Q_2})f_0(\hat{s}),\\
S_1(\hat{s})&=&\frac{1-\hat{m}^2_K}{(\hat{m}_b-\hat{m}_s)}(C_{Q_1}+C'_{Q_1})f_0(\hat{s}),\\
A(\hat{s})&=&\frac{2V(\hat{s})}{1+\hat{m}_{K^*}}
[C_9^{eff}(\hat{s})+C_9^{'eff}(\hat{s})]
+\frac{4\hat{m}_b}{\hat{s}}(C^{eff}_7+C^{'eff}_7)T_1(\hat{s}),\\
B(\hat{s})&=&(1+\hat{m}_{K^*})[C_9^{eff}(\hat{s})-C_9^{'eff}(\hat{s})]A_1(\hat{s})
+ \frac{2\hat{m}_b}{\hat{s}}(1-\hat{m}^2_{K^*})(C^{eff}_7-C^{'eff}_7)T_2(\hat{s}),\\
C(\hat{s})&=&\frac{A_2(\hat{s})}{(1+\hat{m}_{K^*})}[C_9^{eff}(\hat{s})-C_9^{'eff}(\hat{s})]
+\frac{2\hat{m}_b}{1-\hat{m}^2_{K^*}}(C^{eff}_7-C^{'eff}_7)\left(T_3(\hat{s})+\frac{1-\hat{m}^2_{K^*}}{\hat{s}}T_2(\hat{s})\right),\\
E(\hat{s})&=&\frac{2V(\hat{s})}{1+\hat{m}_{K^*}}
(C_{10}+C'_{10}),\\
F(\hat{s})&=&(1+\hat{m}_{K^*})(C_{10}-C'_{10})A_1(\hat{s}),\\
G(\hat{s})&=&\frac{1}{1+\hat{m}_{K^*}}(C_{10}-C'_{10})A_2(\hat{s}),\\
H(\hat{s})&=&\frac{2\hat{m}_{K^*}}{\hat{s}}(C_{10}-C'_{10})\left(A_3(\hat{s})-A_0(\hat{s})\right)
+\frac{\hat{m}_{K^*}}{\hat{m}_l(\hat{m}_b+\hat{m}_s)}(C_{Q_2}-C'_{Q_2})A_0(\hat{s})\\
S_2(\hat{s})&=&\frac{1}{(\hat{m}_b+\hat{m}_s)}A_0(\hat{s})(C'_{Q_1}-C_{Q_1})\;,
\end{eqnarray}
where $f_-(s)=\frac{m_B^2-m_K^2}{s}(f_0(s)-f_+(s)),
A_3(s)=\frac{m_B+m_{K^*}}{2m_{K^*}}A_1(s)-\frac{m_B-m_{K^*}}{2m_{K^*}}A_2(s)
$. The above results reduce to those in ref.~\cite{0004262} if all
$C_i^\prime=0$, as expected. It is worth to note that the final
term in eq. (\ref{nhbs}) vanishes if one does not include the NHB
contributions.

\subsection{The dilepton invarient mass spectra and differential FBA }

The kinematic variables $\hat{s},\hat{u}$ are defined as:
\begin{eqnarray}\label{su}
\hat{s} &=&\hat{q}^2=(\hat{p}_++\hat{p}_-)^2, \nonumber \\
\hat{u}&=&(\hat{p}_B-\hat{p}_-)^2-(\hat{p}_B-\hat{p}_+)^2
\end{eqnarray}

Here we choose the center of mass frame of the dileptons as the
frame of reference, in which the leptons move back to back, and
the momentum of B meson makes an angle $\theta$ with that of
$l^+$. $\hat{u}$ can be written in terms of $\theta$:
\begin{eqnarray}\label{u}
\hat{u}&=&-\hat{u}(\hat{s}) \cdot \cos{\theta}
\equiv -\hat{u}(\hat{s})z, z=\cos{\theta}, \nonumber \\
\hat{u}(\hat{s})&=&\sqrt{\lambda(1-4\frac{\hat{m}^2_l}{\hat{s}})}, {\cal D}=\sqrt{1-4\frac{\hat{m}^2_l}{\hat{s}}},\nonumber \\
\lambda &=&
1+\hat{m}^4_{K^{(*)}}+\hat{s}^2-2\hat{s}-2\hat{m}^2_{K^{(*)}}(1+\hat{s})
\end{eqnarray}

The phase space is defined in terms of $\hat{s}$ and z:
\begin{eqnarray}\label{phase}
(2\hat{m}_l)^2 \leq \hat{s} \leq (1-\hat{m}_{K^{(*)}})^2,-1 \leq z
\leq 1
\end{eqnarray}

   Keeping the lepton mass and integrating over $\hat{u}$ in the kinematic region,
   we can get the dilepton invariant mass
spectra (IMS):
\begin{eqnarray}
\frac{d\Gamma^{K(K^*)}}{d\hat{s}}&=&\frac{G^2_F\alpha^2m^5_B}{2^{10}\pi^5}
|V_{tb}V^*_{ts}|^2 \hat{u}(\hat{s})D^{K(K^*)}, \label{ims}\\
D^K&=&(|A'|^2+|C'|^2)(\lambda-\frac{\hat{u}(\hat{s})^2}{3})
+|S_1|^2(\hat{s}-4\hat{m}^2_l) \nonumber \\
&&+|C'|^2
4\hat{m}^2_l(2+2\hat{m}^2_K-\hat{s})+Re(C'D'^{\dagger})8\hat{m}^2_l(1-\hat{m}^2_K)+|D'|^2
4\hat{m}^2_l\hat{s},\label{ims1}\\
 D^{K^*}&=&
\frac{|A|^2}{3}\hat{s}\lambda(1+2\frac{\hat{m}^2_l}{\hat{s}})
+\frac{|E|^2}{3}\hat{s}\hat{u}(\hat{s})^2
+|S_2|^2(\hat{s}-4\hat{m}^2_l)\lambda \nonumber \\
&&+\frac{1}{4\hat{m}^2_{K^*}}\left[|B|^2(\lambda-\frac{\hat{u}(\hat{s})^2}{3}+8\hat{m}^2_{K^*}(\hat{s}+2\hat{m}^2_l))
+|F|^2(\lambda-\frac{\hat{u}(\hat{s})^2}{3}+8\hat{m}^2_{K^*}(\hat{s}-4\hat{m}^2_l))\right]  \nonumber \\
&&+\frac{\lambda}{4\hat{m}^2_{K^*}}\left[|C|^2(\lambda-\frac{\hat{u}(\hat{s})^2}{3})
+|G|^2\left(\lambda-\frac{\hat{u}(\hat{s})^2}{3}+4\hat{m}^2_l(2+2\hat{m}^2_{K^*}-\hat{s})\right)\right]\nonumber \\
&&-\frac{1}{2\hat{m}^2_{K^*}}\left[Re(BC^{\dagger})(1-\hat{m}^2_{K^*}-\hat{s})(\lambda-\frac{\hat{u}(\hat{s})^2}{3})
\right.\nonumber \\
&&\left.+Re(FG^{\dagger})\left((1-\hat{m}^2_{K^*}-\hat{s})(\lambda-\frac{\hat{u}(\hat{s})^2}{3})
+4\hat{m}^2_l\lambda\right)\right]\nonumber \\
&&-
2\frac{\hat{m}^2_l}{\hat{m}^2_{K^*}}\lambda\left[Re(FH^{\dagger})-Re(GH^{\dagger})(1-\hat{m}^2_{K^*})\right]
+|H|^2\frac{\hat{m}^2_l}{\hat{m}^2_{K^*}}\hat{s}\lambda
\label{ims2}
\end{eqnarray}

 The differential FBA is defined as:
\begin{eqnarray}\label{fba}
 A_{FB(\hat{s})}=
 \frac{-\int^{\hat{u}(\hat{s})}_0dz \frac{d^2\Gamma}{d\hat{s}d\hat{u}}+\int_{-\hat{u}(\hat{s})}^0d\hat{u} \frac{d^2\Gamma}{d\hat{s}d\hat{u}}}
 {\int^{\hat{u}(\hat{s})}_0dz \frac{d^2\Gamma}{d\hat{s}d\hat{u}}+\int_{-\hat{u}(\hat{s})}^0d\hat{u} \frac{d^2\Gamma}{d\hat{s}d\hat{u}}}
\end{eqnarray}
According to the definition, it is straightforward to obtain the
expressions of FBA in the exclusive decays:
\begin{itemize}
\item $B \to K l^+l^-$
\begin{eqnarray}\label{fbak}
\frac{dA^K_{FB}}{d\hat{s}}D^K=-2\hat{m}_l\hat{u}(\hat{s})Re(S_1A'^{\dagger})
\end{eqnarray}
\item $B \to K^* l^+l^-$
\begin{eqnarray}\label{fbaks}
\frac{dA^{K^*}_{FB}}{d\hat{s}}D^{K^*}=\hat{u}(\hat{s})\left\{\hat{s}[Re(BE^{\dagger})+Re(AF^{\dagger})]
+\frac{\hat{m}_l}{\hat{m}_{K^*}}[Re(S_2B^{\dagger})(1-\hat{s}-\hat{m}^2_{K^*})-Re(S_2C^{\dagger})\lambda]\right\}
\end{eqnarray}
\end{itemize}

   Seen from the Eqs. (\ref{ims}), (\ref{ims1}), (\ref{ims2}), (\ref{fbak}) and (\ref{fbaks}), the functions
$D'(\hat{s}),S_1(\hat{s}),H(\hat{s})$, and $S_2(\hat{s})$, which
come from the contribution of NHBs, enter the IMS and FBA. Hence,
the effects of NHBs will manifest themselves in the numerical
results of these formula. In particular, Eq. (\ref{fbak}) shows
that the FBA in $B\to K l^+l^-$ vanishes if there is no NHB
contributions and from Eq. (\ref{fbaks}) it follows that the NHB
contributions change the position of the zero-point of the FBA in
$B\to K^* l^+l^-$. As pointed out in ref.~\cite{0209228}, in an
untagged sample, the FB asymmetry for unpolarized leptons
vanishes. Once the flavor of the decaying b-quark is tagged, one
can measure the unpolarized FB asymmetry which is an important
observable to discriminate new physics from the SM, as we noted
above.
\subsection{The lepton polarization}

   In this subsection, we will present the analytical expressions of lepton
polarization. We define the three orthogonal unit vectors in the
center mass frame of dilepton as
\begin{eqnarray}
      \hat{e}_L&=&\vec{p}_+,\nonumber \\
      \hat{e}_N&=&\frac{\vec{p}_K \times \vec{p}_+ }{|\vec{p}_K \times \vec{p}_+|},\nonumber \\
      \hat{e}_T&=& \hat{e}_N \times \hat{e}_L\; ,
\end{eqnarray}
which are related to the spin of lepton by a Lorentz boost.
  Then, the decay width of the $B \to K^{(*)} l^+ l^-$ decay for any spin direction
$\hat{n}$ of the lepton, where $\hat{n}$ is a unit vector in the
dilepton center mass frame, can be written as:
\begin{eqnarray}
      \frac{d\Gamma(\hat{n})}{d\hat{s}}=\frac{1}{2}\big (\frac{d\Gamma}{d\hat{s}}\big )_0[1
      +(P_L\hat{e}_L+P_N\hat{e}_N+P_T\hat{e}_T)\cdot\hat{n}]
\end{eqnarray}
where the subscript $"0"$ denotes the unpolarized decay width,
$P_L$ and $P_T$  are the longitudinal and transverse polarization
asymmetries in the decay plane respectively, and $P_N$ is the
normal polarization asymmetry in the direction perpendicular to
the decay plane.

  The lepton polarization asymmetry $P_i$ can be obtained by calculating
\begin{eqnarray}
P_i(\hat{s})=\frac{d\Gamma(\hat{n}=\hat{e}_i)/d\hat{s}-
d\Gamma(\hat{n}=-\hat{e}_i)/d\hat{s}}{d\Gamma(\hat{n}=\hat{e}_i)/d\hat{s}+
d\Gamma(\hat{n}=-\hat{e}_i)/d\hat{s}}\; .
\end{eqnarray}
By a straightforward calculation, we get
\begin{itemize}
\item $B \to K l^+l^-$
\begin{eqnarray}
P^K_LD^K&=&\frac{4}{3}{\cal D}\left\{\lambda
Re(A'C'^{\dagger})-3\hat{m}_l(1-\hat{m}^2_{K})Re(C'^{\dagger}S_1)-3\hat{m}_l\hat{s}Re(D'^{\dagger}S_1)\right\}, \\
P^K_ND^K&=&\frac{\pi\sqrt{\hat{s}}\hat{u}(\hat{s})}{2}\left\{-Im(A'S_1^{\dagger})+2\hat{m}_lIm(C'D'^{\dagger})\right\},\label{pnk} \\
P^K_TD^K&=&-\frac{\pi\sqrt{\lambda}}{\sqrt{\hat{s}}}\left\{\hat{m}_l\left[(1-\hat{m}^2_{K})Re(A'C'^{\dagger})+\hat{s}Re(A'D'^{\dagger})\right]
+\frac{(\hat{s}-4\hat{m}^2_l)}{2}Re(C'S_1^{\dagger})\right\}
\end{eqnarray}
\item $B \to K^*l^+l^-$
 \begin{eqnarray}
 P^{K^*}_LD^{K^*}&=&{\cal D}\left\{\frac{2\hat{s}\lambda}{3}
 Re(AE^{\dagger})+\frac{(\lambda+12\hat{m}^2_{K^*}\hat{s})}{3\hat{m}^2_{K^*}}Re(BF^{\dagger})\right.\nonumber\\
 &&\left.-\frac{\lambda(1-\hat{m}^2_{K^*}-\hat{s})}{3\hat{m}^2_{K^*}}Re(BG^{\dagger}+CF^{\dagger})
 +\frac{\lambda^2}{3\hat{m}_{K^*}}Re(CG^{\dagger})\right.\nonumber\\
 &&\left.+\frac{2\hat{m}_l\lambda}{\hat{m}_{K^*}}[Re(FS^{\dagger}_2)-\hat{s}Re(HS^{\dagger}_2)
 -(1-\hat{m}^2_{K^*})Re(GS^{\dagger}_2)]\right\} ,\\
 P^{K^*}_ND^{K^*}&=&\frac{-\pi\sqrt{\hat{s}}\hat{u}(\hat{s})}{4\hat{m}_{K^*}}\left\{
 \frac{\hat{m}_l}{\hat{m}_{K^*}}\left[Im(FG^{\dagger})(1+3\hat{m}^2_{K^*}-\hat{s})\right.\right. \nonumber\\
 &&\left.\left.+Im(FH^{\dagger})(1-\hat{m}^2_{K^*}-\hat{s})-Im(GH^{\dagger})\lambda  \right]\right. \nonumber\\
 &&\left.+2\hat{m}_{K^*}\hat{m}_l[Im(BE^{\dagger})+Im(AF^{\dagger})]\right.\nonumber\\
 &&\left.-(1-\hat{m}^2_{K^*}-\hat{s})Im(BS^{\dagger}_2)+\lambda Im(CS_2^{\dagger})\right\},\\
P^{K^*}_TD^{K^*}&=&\frac{\pi\sqrt{\lambda}\hat{m}_l}{4\sqrt{\hat{s}}}\left\{ 4\hat{s}Re(AB^{\dagger})\right.\nonumber\\
&&\left.+\frac{(1-\hat{m}^2_{K^*}-\hat{s})}{\hat{m}^2_{K^*}}\left[
-Re(BF^{\dagger})+(1-\hat{m}^2_{K^*})Re(BG^{\dagger})+\hat{s}Re(BH^{\dagger})\right]\right.\nonumber\\
&&\left.+\frac{\lambda}{\hat{m}^2_{K^*}}[Re(CF^{\dagger})-(1-\hat{m}^2_{K^*})Re(CG^{\dagger})
-\hat{s}Re(CH^{\dagger})]\right.\nonumber\\
&&\left.+\frac{(\hat{s}-4\hat{m}^2_l)}{\hat{m}_{K^*}\hat{m}_l}
[(1-\hat{m}^2_{K^*}-\hat{s})Re(FS^{\dagger}_2)-\lambda
Re(GS^{\dagger}_2)]\right\}
\end{eqnarray}
\end{itemize}

One can see from Eq. (\ref{pnk}) that $P_N=0$ for the decay $B\to
K l^+l^-$ in the SM because $C_{10}^\prime=0$ in the approximation
of $m_s/m_b=0$, $C_{Q_{1,2}}^{(\prime)}=0$, and $C_{10}$ is real
in the SM. Thus, a non zero normal polarization asymmetry in $B\to
K l^+l^-$ would signal the existence of new physics.

\section{Mass spectra and the permitted parameter space}

To see the impact of the induced off-diagonal elements in the mass
matrix of the right-handed down-type squarks on B rare decays and
simplify the analysis, we assume that at the GUT scale ($M_G$) all
sfermion mass matrices except the right-handed down-type squark
mass matrix are flavor diagonal and all diagonal elements are
approximately universal and equal to $m^2_0$. The 2-3 matrix
element of right-handed down-type squark mass matrix is
parameterized by $\delta_{23}^{dRR}\equiv
\frac{(M^2_{\tilde{d}RR})_{23}}{m_0^2}$ which can be treated as a
free parameter of order one. Furthermore, we have a universal
gaugino mass $M_{1/2}$, a universal trilinear coupling $A_0$ and a
universal bilinear coupling $B_0$ at $M_G$. Taking account to the
radiative electro-weak (EW) symmetry breaking, finally we have
five parameters ($m_0, M_{1/2}, A_0, \delta_{23}^{dRR}, tan\beta$)
plus the sign of $\mu$ as the initial conditions for solving the
renormalization group equations (RGEs).

 We require the lightest neutralino to be the lightest
supersymmetric particle (LSP) and use several experimental limits
to constraint the parameter space, including 1)the width of the
decay $Z \rightarrow \chi^0_1 \chi^0_1$ is less than 4.3 MeV, and
branching ratios of $Z \rightarrow \chi^0_1 \chi^0_2$ and $Z
\rightarrow \chi^0_2 \chi^0_2$ are less than $1 \times 10^{-5}$,
where $\chi^0_1$ is the lightest neutralino and $\chi^0_2$ is the
other neutralino; 2) the mass of light neutral Higgs can not be
lower than 111 GeV as the present experments required; 3) the mass
of lighter chargino must be larger than 94 GeV as given by the
Particle Data Group~\cite{pdg}; 4) sneutrinos are larger than
94GeV; 5) seletrons are larger than 73GeV; 6) smuons larger than
94 GeV; 7) staus larger than 81.9 GeV.

 In the numerical calculation, we use the revised ISAJET.  We find
that the parameter $\delta_{23}^{dRR}$ does not receive any
significant correction and the diagonal entries of mass matrices
are significantly corrected, which is in agreement with the
results in Ref.~\cite{cmsvv}. We scan $m_0,\;M_{1/2}$ in the range
(100, 800) GeV for given values of $A_0, \tan\beta$ and
sign($\mu$)=+1\footnote{In the case of sign($\mu$)= -1, the
constraint from $B\to X_s \gamma$ on the parameter space is too
stringent, in particular, for large
$\tan\beta$~\cite{nanop,dyb,hy}}, with the constraints from the
relevant low energy experiments such as $B \to X_s \gamma$,
$B_s\to \mu^+\mu^-$, etc. (for the detailed discussions of
constraints, see VB).

\begin{figure}
{\includegraphics[width=8.3cm] {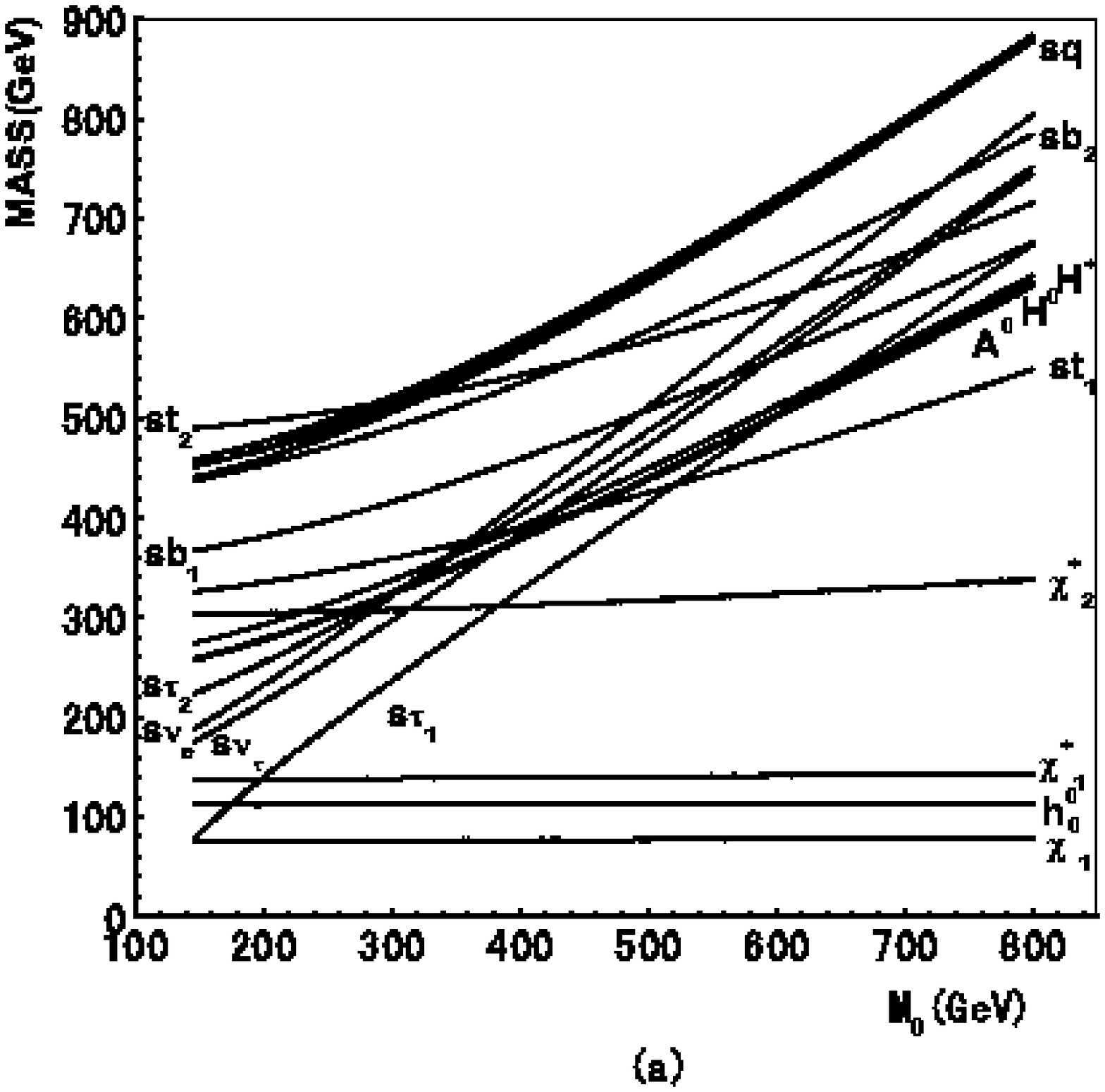}}
{\includegraphics[width=8.6cm] {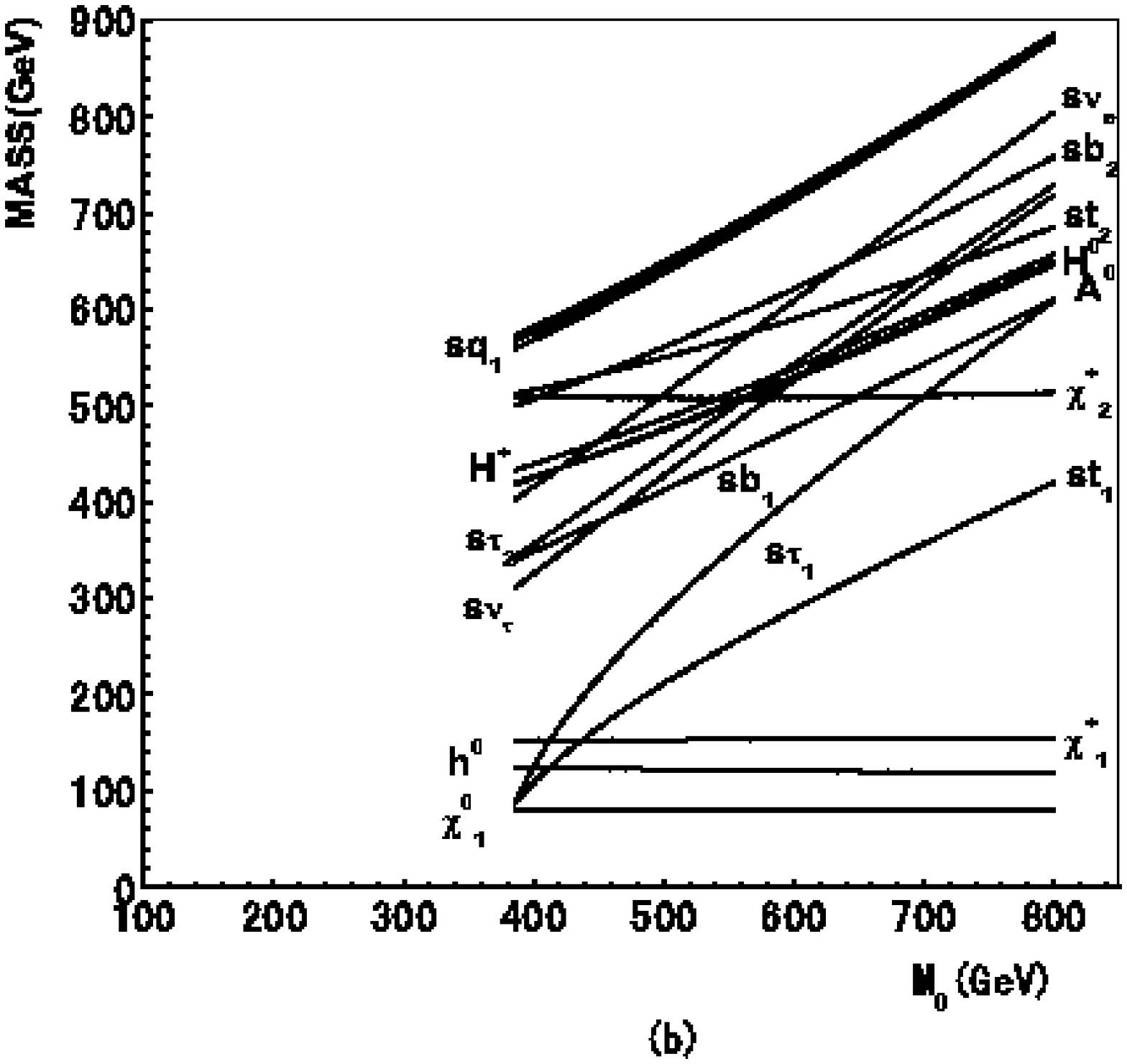}} \caption{
\label{noncon} The mass spectrum versus $m_0$ for fixed
$M_{1/2}$=200 Gev, $tan\beta$=40, $\delta^d_{23RR}$=(0.04-0.03i),
and sign($\mu$)=+1 without the constraints from the low energy
experiments imposed. (a) is for $A_0=0$. (b) is for $A_0$=-1000.}
\end{figure}

\begin{figure}
{\includegraphics[width=8.8cm] {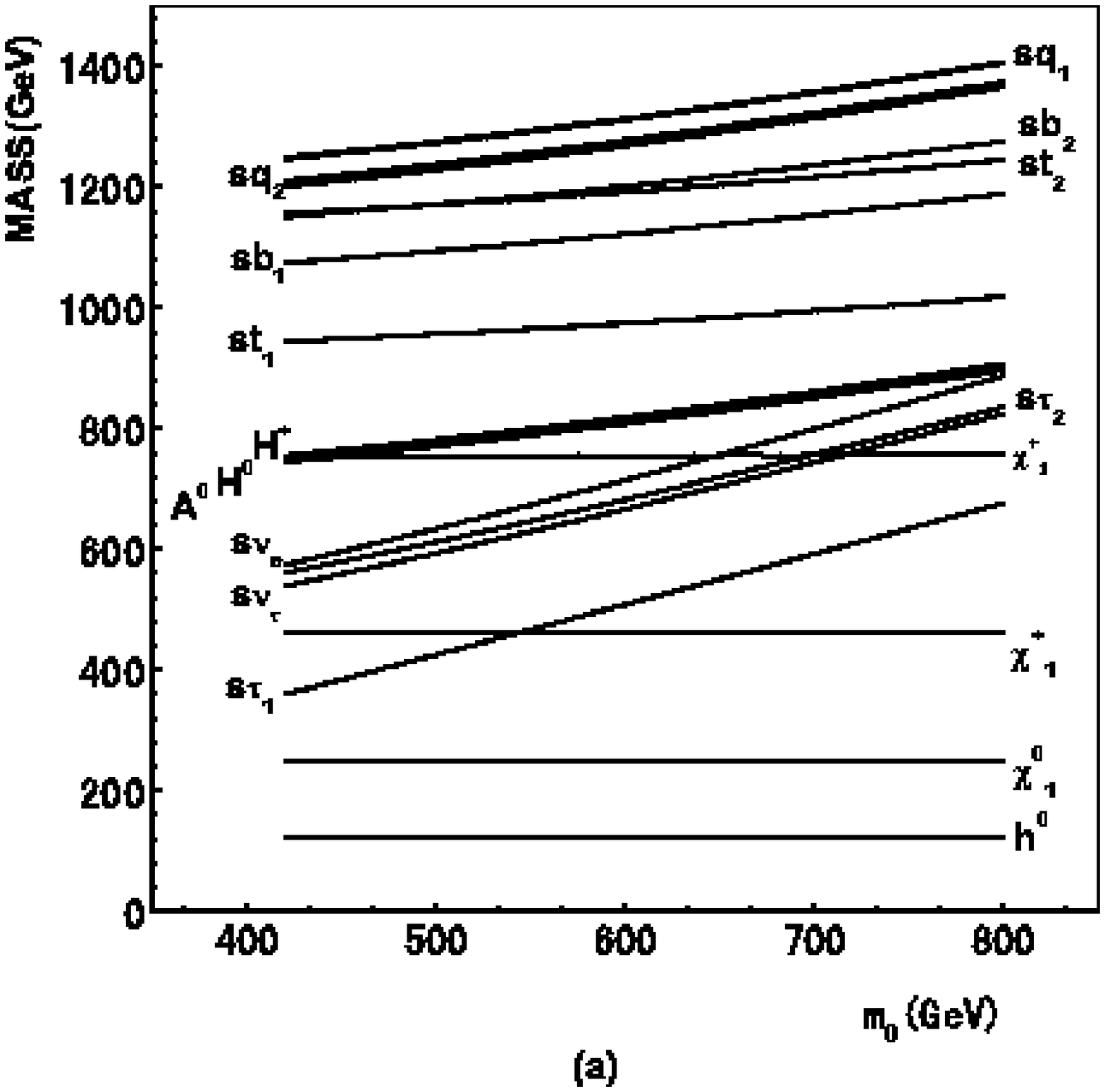}}
{\includegraphics[width=8.8cm] {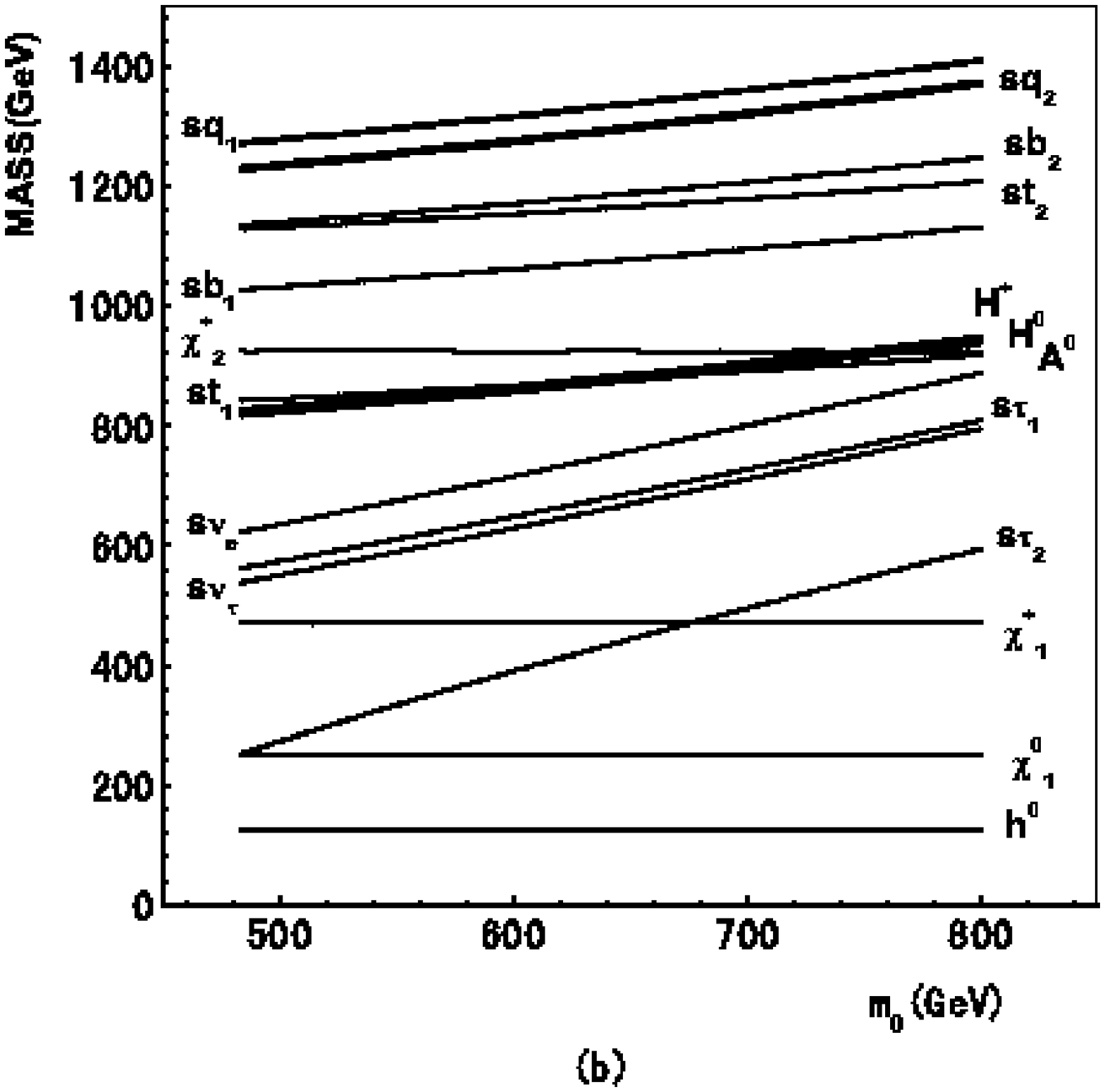}} \caption{
\label{con} The mass spectrum versus $m_0$ for fixed $M_{1/2}$=600
Gev, $tan\beta$=40, $\delta^d_{23RR}$=(0.04-0.03i), and
sign($\mu$)=+1 with the constraints from the low energy
experiments imposed. (a) is for $A_0=0$. (b) is for $A_0$=-1000.}
\end{figure}

   For an illustration, we present the mass spectra without and with the
constraints from the low energy experiments in {\bf Figs}. 1 and
2, respectively, where (a) and (b) are for $A_0=0, -1000 GeV$
respectively. One can see from the {\bf Figs}. \ref{noncon},
\ref{con} that the mass spectrum lifts when $A_0$ increases and
when the constraints from the low energy experiments are imposed
the masses of sparticles are larger than those without the
constraints, as expected.

\section{Numerical analysis}

In this section, we will discuss the numerical results and make an
     analysis.
\subsection{Parameters input}
      Parameters in our calculation are list :
\begin{eqnarray}
\begin{array}{ll@{\,}@{\qquad}lll}
&m_b = 4.8 GeV , m_c= 1.4 GeV ,&m_s= 0.2 GeV ,m_{\mu}=0.1057 GeV, &m_{\tau} = 1.7769 GeV, \nonumber\\
&M_B = 5.28 GeV,M_{J/\psi} = 3.10 GeV ,&M_{\psi'} = 3.69 GeV,M_{K^*}= 0.89 GeV,&M_K =0.49 GeV ,\nonumber\\
&\Gamma_B=4.22 \times 10^{-13} GeV , \  \  &\Gamma_{J/\psi}=8.70
\times 10^{-5}GeV , \ &\Gamma_{\psi'}=27.70 \times
10^{-5}GeV  \nonumber \\
& \Gamma({J/\psi} \to l^+l^-)=5.26 \times 10^{-6}GeV , \
&\Gamma({\psi'} \to l^+l^-)=2.14 \times 10^{-6}GeV
\end{array}
\end{eqnarray}
\subsection{Constraints from experiments}

      In our calculation, we consider the constraints from
$B \to X_s \gamma$, $B_s \to \mu^+ \mu^-$, $\Delta M_s$, $B \to
K^{(*)} l^+l^-$, and $\tau\to\mu\gamma$. The leading order $B_s
\to X_s \gamma$ branching ratio normalized to $Br(B \to  X_c e
\bar{\nu})$ can be written as
$$
Br(B \to  X_s \gamma ) = \frac{6\alpha_{em} }{\pi f(z) }|
\frac{V_{tb}V_{ts}^*}{ V_{cb}}|^2 Br(B \to  X_c e
\bar{\nu})(|C_7(m_b)|^2 + |C'_7(m_b)|^2)$$
 where $\sqrt{z} = m_c^{pole}/m_b^{pole}$ , $f(z)$ is the phase space function.
We take $2 \times 10^{-4} < Br(B \to X_s \gamma) < 4.5 \times
10^{-4}$, considering the theoretical uncertainties. The $B \to
X_s \gamma$ make a direct constraint on $|C_7(m_b)|^2 +
|C'_7(m_b)|^2$. The single insertion term $\delta_{23}^{LR(RL)}$
in $C'_7(m_b)$ are more severely constrained than
$\delta_{23}^{LL(RR)}$, due to the strong enhancement factor
$m_{\tilde{g}}/m_b$ associated with single $\delta^{dLR(RL)}_{23}$
insertion term in $C^{(\prime)}_{7}(m_b)$. Because the double
insertion term $\delta^{dLL(RR)}_{23} \delta^{dLR(LR*)}_{33}$ is
also enhanced by $m_{\tilde{g}}/m_b$, $\delta^{dLL(RR)}_{23}$ is
constrained to be order of $10^{-2}$ if the left-right mixing of
scalar bottom quark $\delta^{dLR(RL)}_{33}$ is large ($\sim 0.5$).
Nevertheless, in the large $\tan\beta$ case the chargino
contribution can destructively interfere with the SM (plus the
charged Higgs) contribution so that the constraint can be easily
satisfied.

   The branching ratio $Br(B \to \mu^+ \mu^-)$ is given as~\cite{wuxh}
\begin{eqnarray}\label{bsmu}
{\rm Br}(B_s \rightarrow \mu^+ \mu^-) &=& \frac{G_F^2
\alpha^2_{\rm em}}{64 \pi^3} m^3_{B_s} \tau_{B_s} f^2_{B_s}
|\lambda_t|^2 \sqrt{1 - 4 \widehat{m}^2}
[(1 - 4\widehat{m}^2) |C_{Q_1}(m_b) - C^\prime_{Q_1}(m_b)|^2 + \nonumber\\
&& |C_{Q_2}(m_b) - C^\prime_{Q_2}(m_b) + 2\widehat{m}(C_{10}(m_b)
- C^\prime_{10}(m_b) )|^2]
\end{eqnarray}
where $\widehat{m} = m_{\mu}/m_{B_s}$. With large
$C^{(\prime)}_{Q_{1,2}}$, $Br(B \to \mu^+ \mu^-)$ can have large
enhancements~\cite{bsmu}. The new $D_0$ experimental upper bound
of ${\rm
Br}(B_s\to \mu^+\mu^-)$ is $4.6\times 10^{-7}$ 
~\cite{d0} at $90\%$ confidence level. It gives a stringent
constraint on $C^{(\prime)}_{Q_{1,2}}$ and consequently the
parameter space of the model. At the same time we require that
predicted branching ratios of $B\to X_s \mu^+\mu^-$ and $B\to
K^{(*)} \mu^+\mu^-$ falls within 1 $\sigma$ experimental bounds.

We also impose the current experimental lower bound $\Delta M_s
> 14.4 ps^{-1}$~\cite{msd}. The $\delta^{dLR(RL)}_{23}$
contribution to $\Delta M_s$ is small because it is constrained to
be order of $10^{-2}$ by Br($B \to X_s \gamma$). The dominant
contribution to $\Delta M_s$ comes from $\delta^{dLL(RR)}_{23}$
insertion with both constructive and destructive effects compared
with the SM contribution, where the too large destructive effect
is ruled out, because SM prediction is only slightly above the
present experiment lower bound.

Furthermore, as analyzed in ref.~\cite{cmsvv}, there is the
correlation between flavor changing squark and slepton mass
insertions in SUSY GUTs. The correlation leads to a bound on
$\delta_{23}^{dRR}$ from the rare decay $\tau\to\mu\gamma$. We
update the analysis with latest BELLE upper bound of
Br($\tau\to\mu\gamma$)$ < 3.1 \times 10^{-7}$~\cite{taumugamma} at
$90\%$ confidence level in the SUSY SO(10) model.
\begin{table}[t]\label{cq12}
\caption{The Wilson coefficients for the two cases in the $SUSY
SO(10)$. The SM values also are listed for comparison. The values
in bracket are for $l=\tau$.}
\begin{tabular}{|c|c|c|c|c|c|}
\hline
$A_0$&$C_{Q_{1}}$ &$C_{Q_{1}}^\prime$&$C_{Q_{2}}$&$C_{Q_2}^\prime$   \\
\hline SM&0&0&0&0 \\
\hline $A_0=0$&0.074+0.I(1.252+0.001I)&-0.013+0.008I(-0.213+0.128I)&-0.075+0.I(-1.267-0.001I)&-0.013+0.008I(-0.216+0.129I)\\
\hline$A_0=-1000
$&0.106+0.I(1.775+0.002I)&-0.247+0.242I(-4.148+4.074I)&-0.107+0.I(-1.797-0.002I)&-0.250+0.246I(-4.202+4.128I)
\\\hline
\end{tabular}
\begin{tabular}{|c|c|c|c|c|c|c|c|}
\hline $A_0$&$C_7^{eff}$&$C_7^{'eff}$&$C_9$&$C'_9$&$C_{10}$&$C'_{10}$ \\
\hline SM&-0.313&0&+4.344&0&-4.669&0 \\
\hline
$A_0=0$&-0.225-0.I&-0.02-0.01I&4.277+0.I&-0.-0.002I&-4.717-0.I
&-0.001+0.019I \\
\hline$A_0=-1000$&-0.219+0.I&0.039-0.038I&4.275+0.I
&0.011+0.072I&-4.732-0.I& -0.075-0.670I
\\\hline
\end{tabular}
\end{table}

\subsection{ The Numerical results and Conclusions}

 We will focus on the parameter space at large $tan\beta$. The reason
for this is that in the large $tan\beta$ region of parameter space
the contributions of NHB exchange become very important for quark
level semi-leptonic transitions $b \to s l^+l^-$ when the final
state lepton is either a muon or tau~\cite{dyb,hy}. In numerical
calculations, we take $\tan\beta$=40, sign($\mu$)=+1, and
$A_0=0,-1000$ and get the sparticle mass spectrum and mixing at
the EW scale. Using the resulted Wilson coefficients, we calculate
the IMS, FBA and polarization asymmetries of processes $ B \to
K^{(*)} l^+l^-$ under the constraints on $\delta_{23}^{dRR}$ from
the all relevant experiments as discussed in subsection VB, whose
phase varies from 0 to $2\pi$.

  The Wilson coefficients $C_{Q_1}^{(\prime)}$ and  $C_{Q_2}^{(\prime)}$
come from NHB exchanging. Specially, we are interested in the case
of maximal enhancements of $C_{Q_{1,2}}^{(\prime)}$. Through
scanning the parameter space under constraints, we find, for
$A_0=0$, when $m_0$=800GeV, $M_{1/2}$=400GeV (for $A_0$=-1000,
$M_{1/2}$=500GeV), $C_{Q_{1,2}}^{(\prime)}$ have maximal values.
The obtained $C_{Q_{1,2}}^{(\prime)}$ and other relevant Wilson
coefficients in the two cases and in SM are listed in Table I.
From Table I, we can know: (i) the NHB contributions in the case
of $A_0=-1000$ are larger than those in the case of $A_0=0$ except
for $C_9$; (ii) the Wilson coefficients $C_{Q_{1,2}}^\prime$ of
primed operators are dominant, which is due to the presence of
$\delta_{23}^{dRR}$ of order one at the high scale in the SUSY
SO(10) model, and their imaginary parts are sizable, which
contribute to the normal polarization, a T violating observable.

\begin{figure}
{\includegraphics[width=8.7cm] {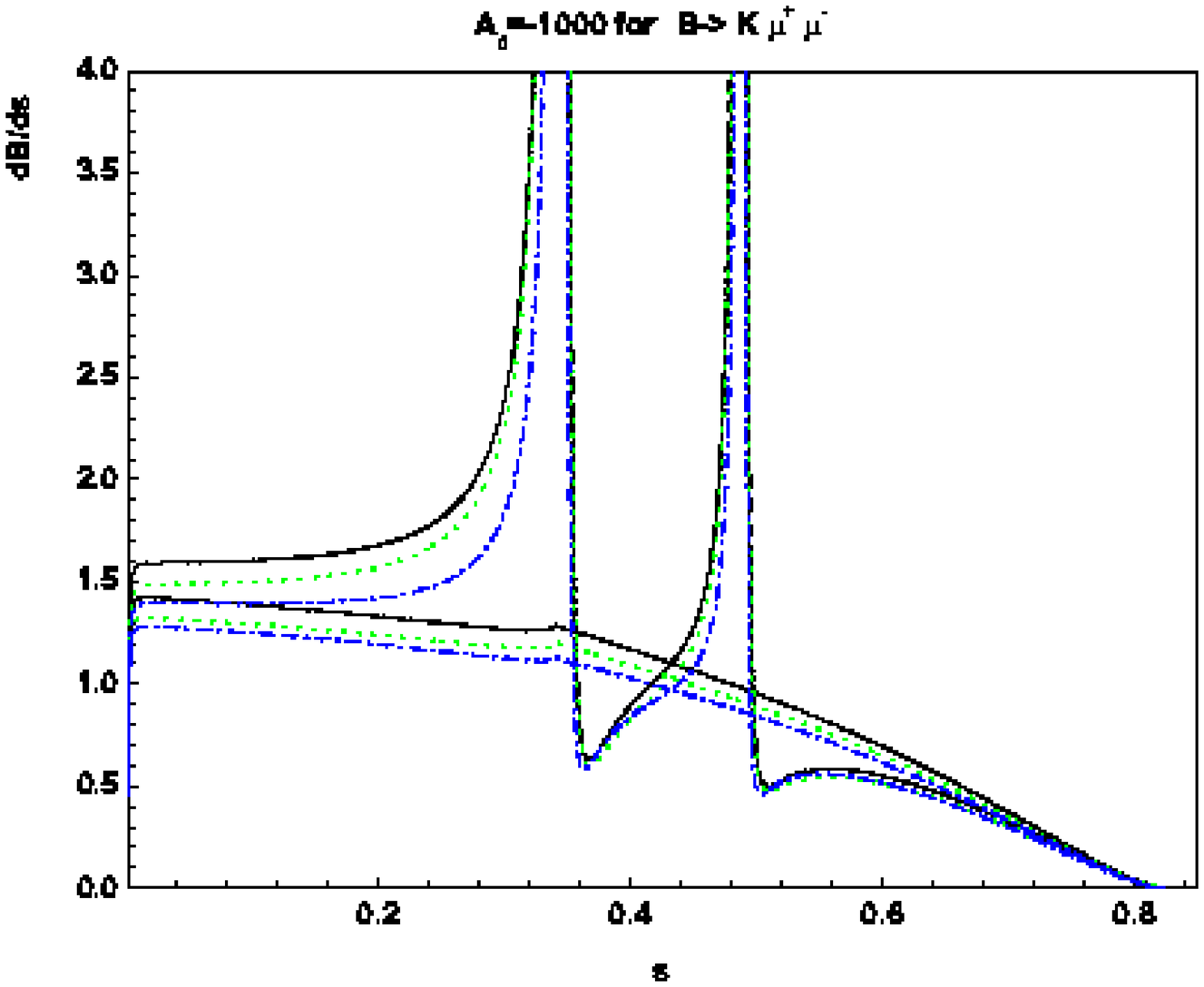}}
{\includegraphics[width=8.9cm] {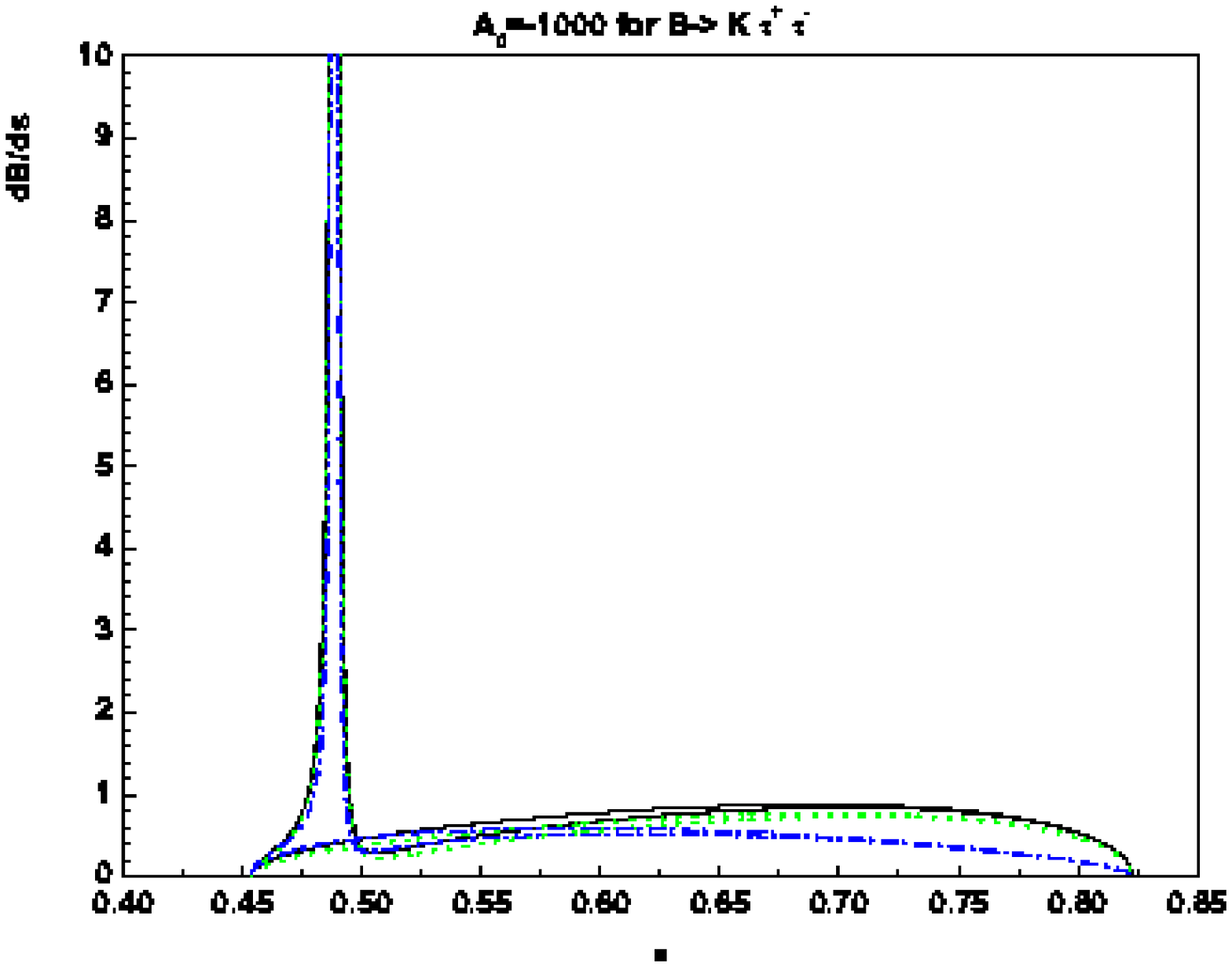}}
{\includegraphics[width=8.9cm] {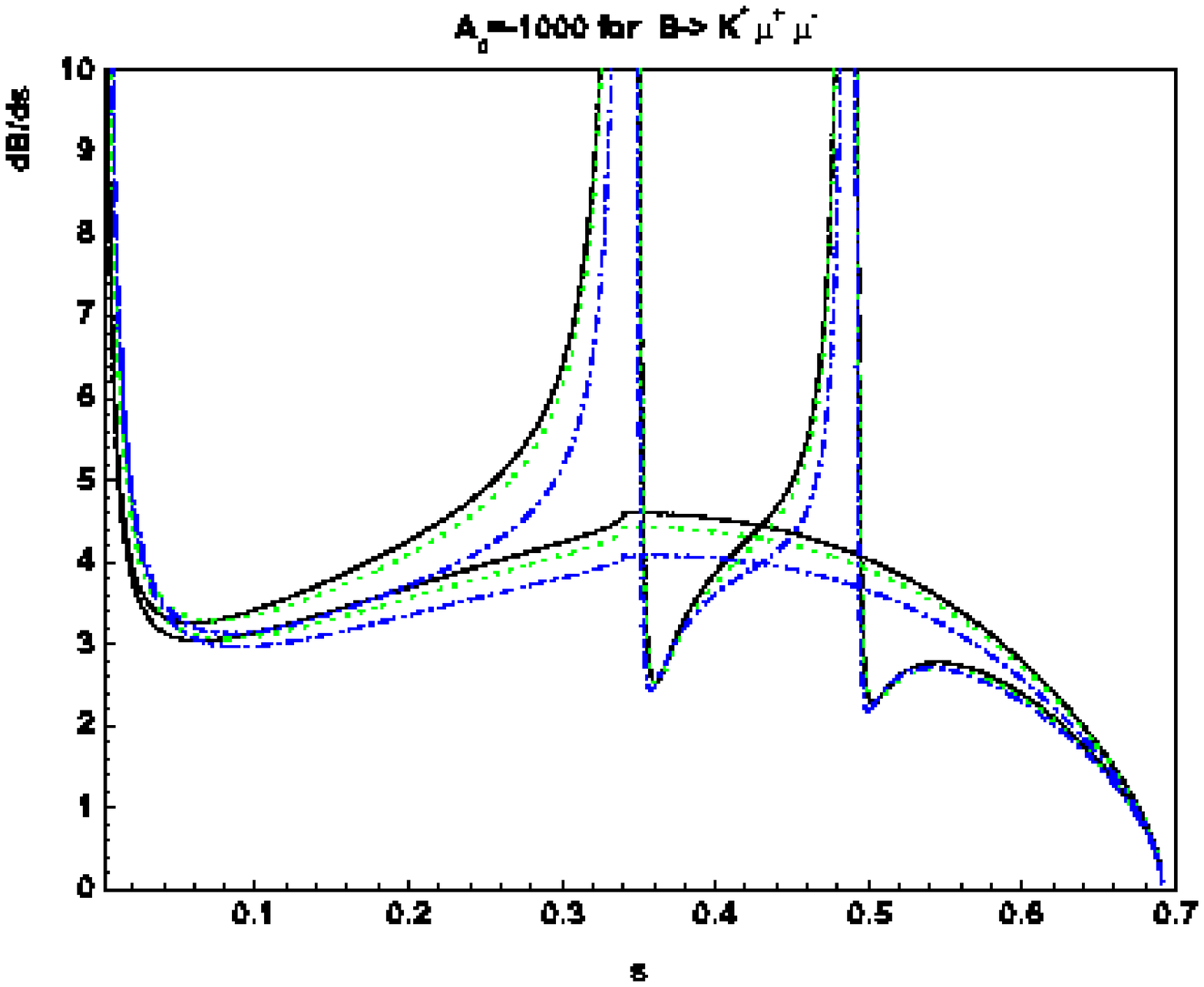}}
{\includegraphics[width=8.9cm] {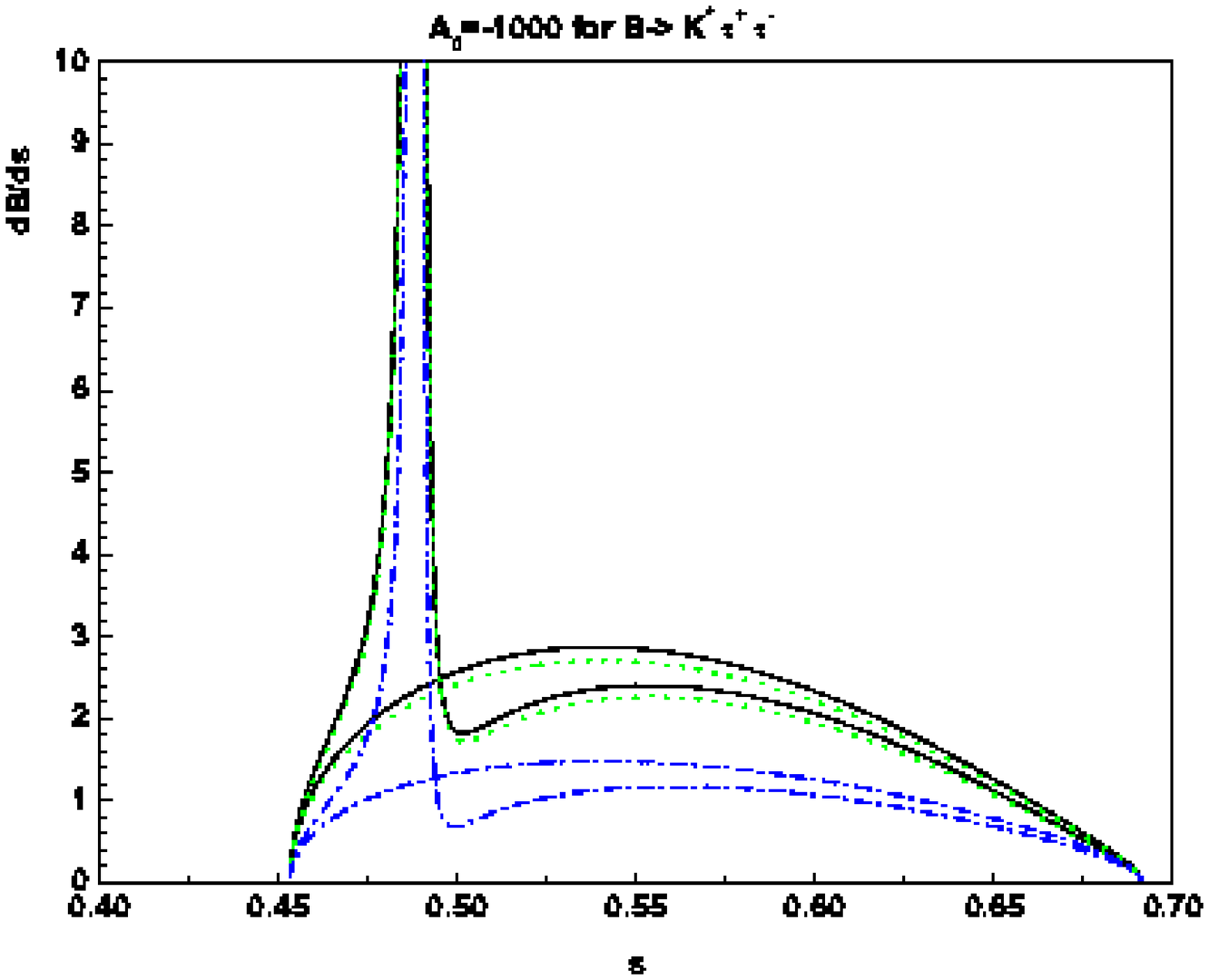}} \caption{
\label{bkims} The IMS of the process $B \to K^{(*)} l^+ l^-$ for
$A_0= -1000$. The solid line(black), dot line(green), and
dashed-dot line(blue) represent the all contributions included,
the SM contributions plus only the NHB contributions, and the SM
contributions, respectively. Both the total (SD+LD) and the pure
SD contributions are shown in order to compare. In the fig. we
write "s" in stead of "$\hat{s}$" for simplicity.}
\end{figure}

\begin{figure}
{\includegraphics[width=8.7cm] {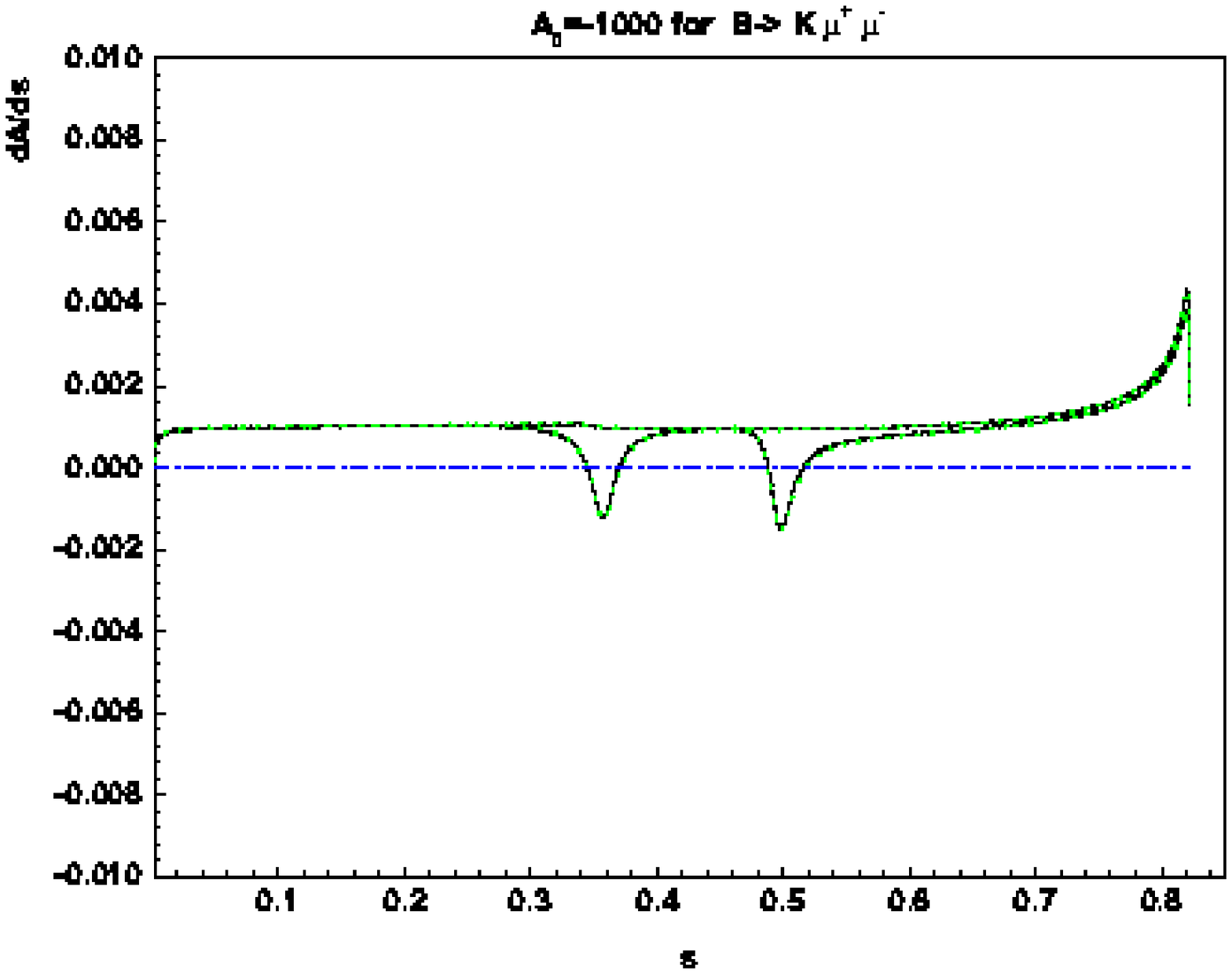}}
{\includegraphics[width=8.9cm] {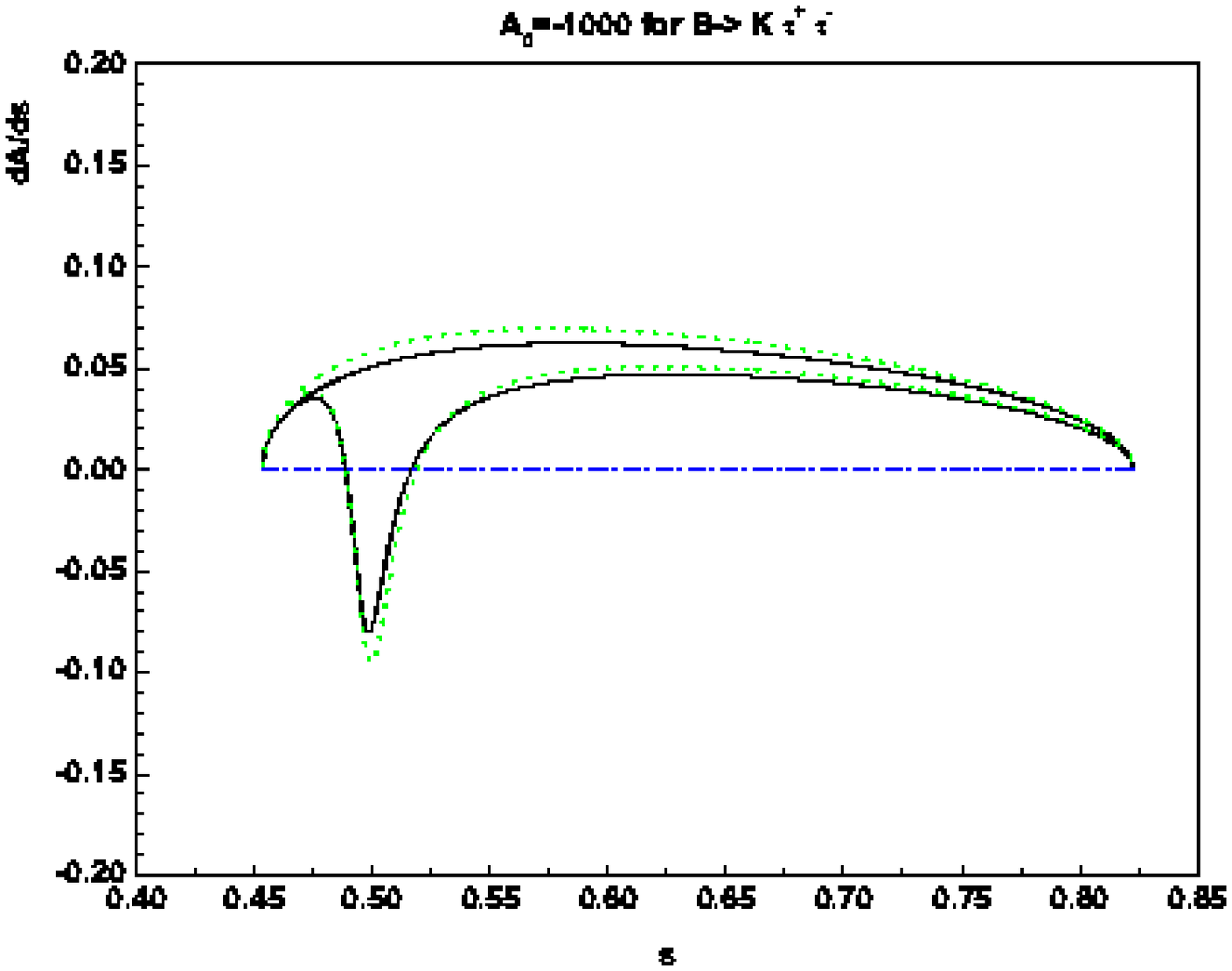}}
{\includegraphics[width=8.9cm] {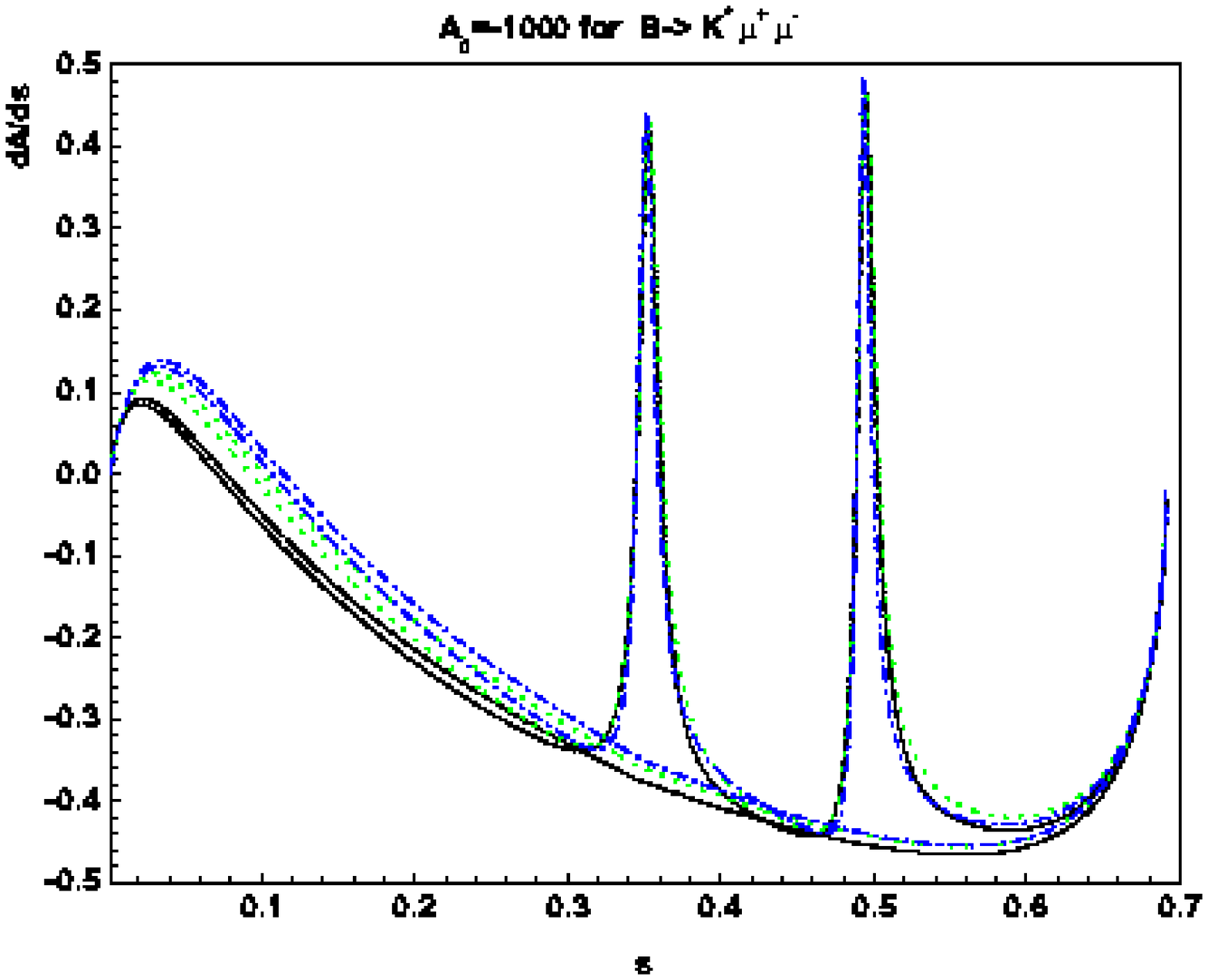}}
{\includegraphics[width=8.9cm] {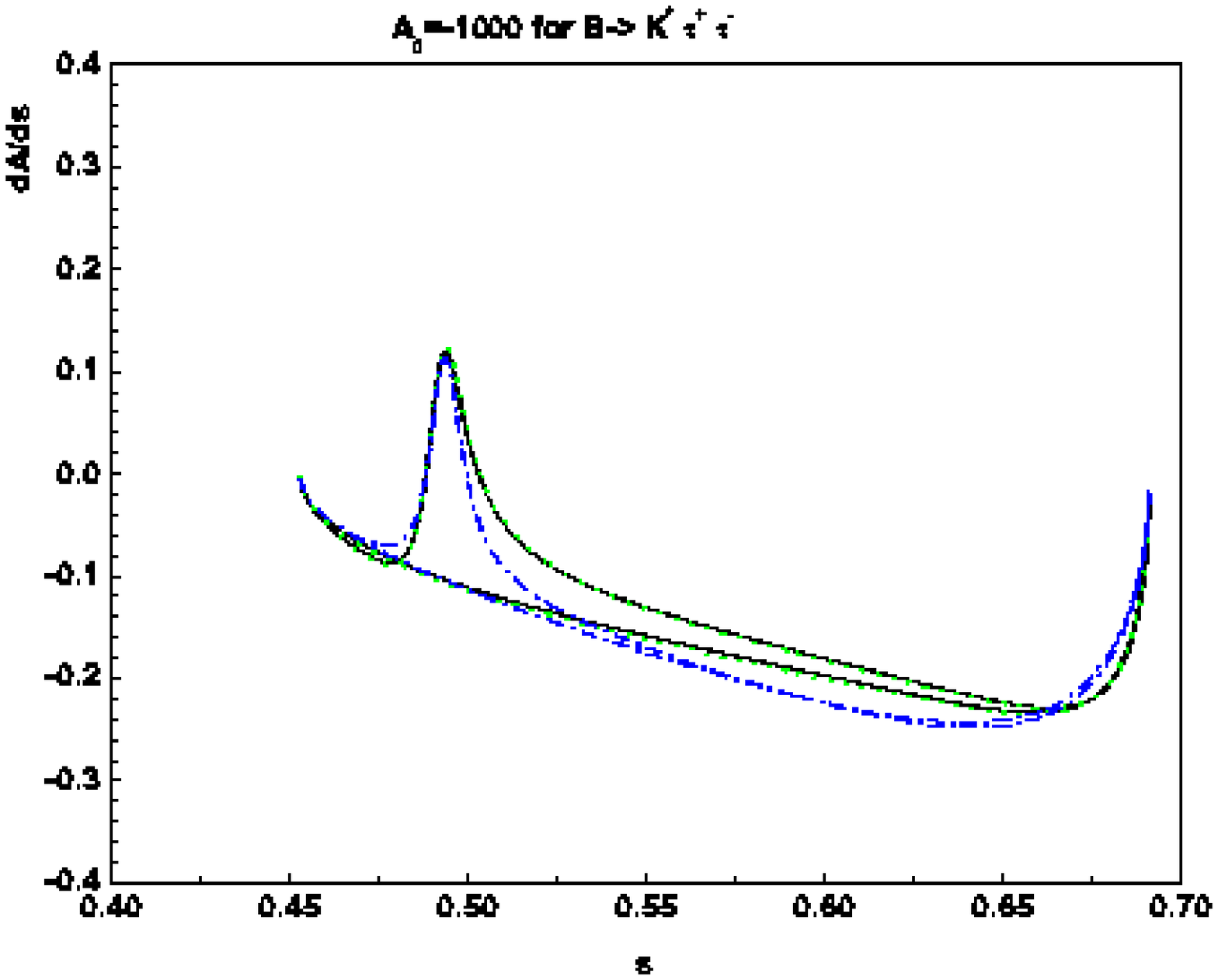}} \caption{
\label{bkfba} The FBA of the process $B \to K^{(*)} l^+ l^-$ for
$A_0= -1000$. The line conventions are the same as those in Fig.
\ref{bkims}.}\end{figure}

\begin{figure}
{\includegraphics[width=8.7cm] {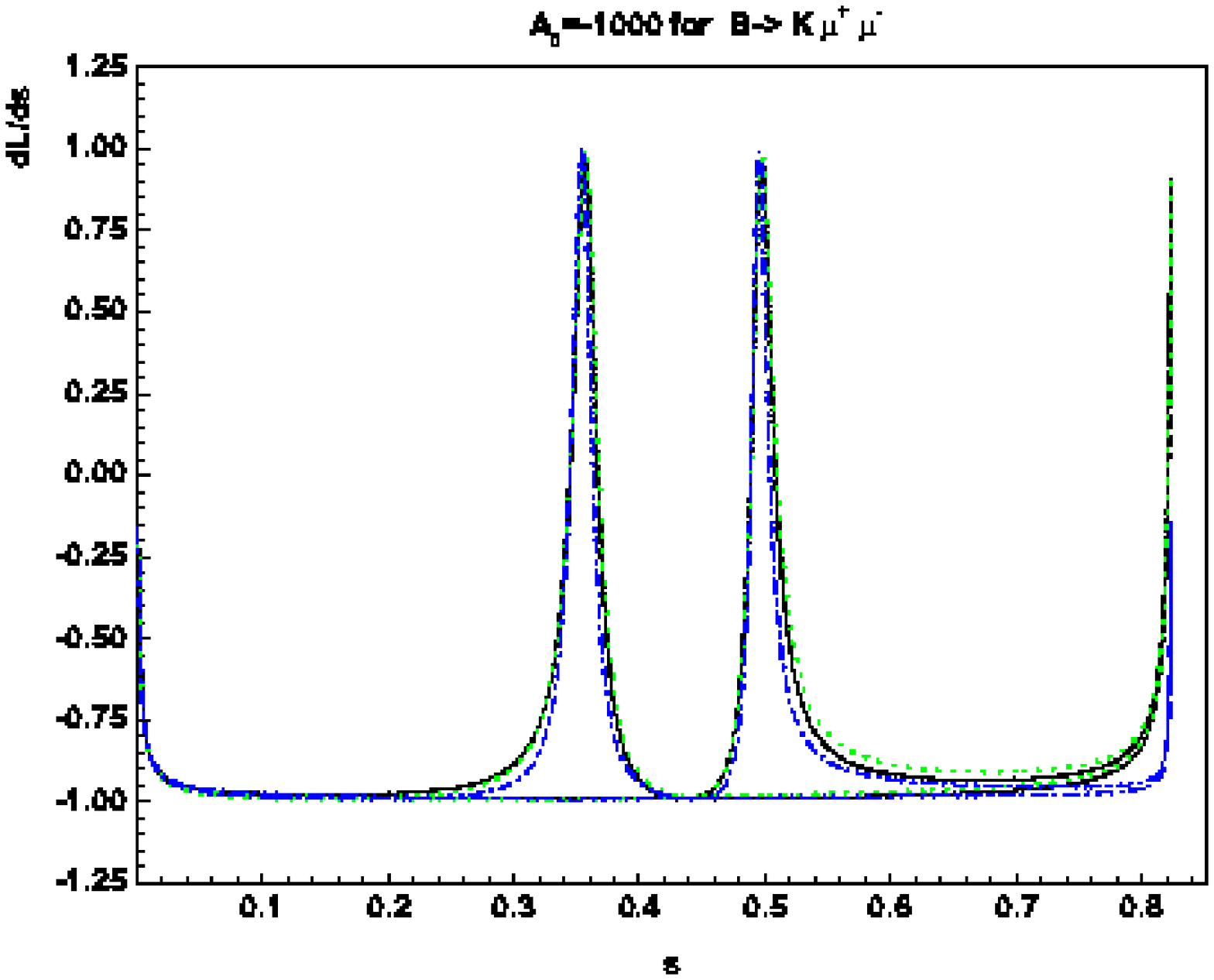}}
{\includegraphics[width=8.9cm] {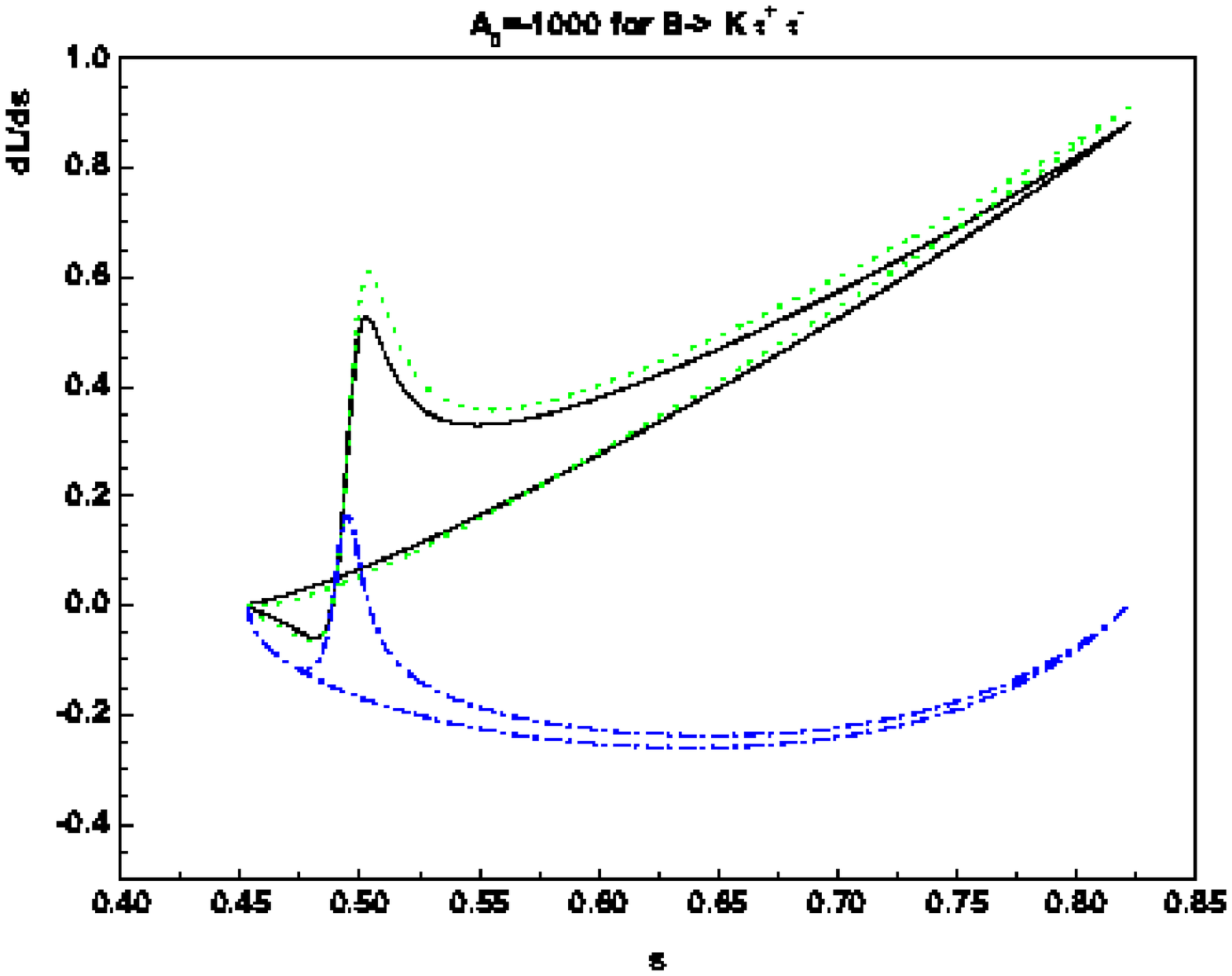}}
{\includegraphics[width=8.9cm] {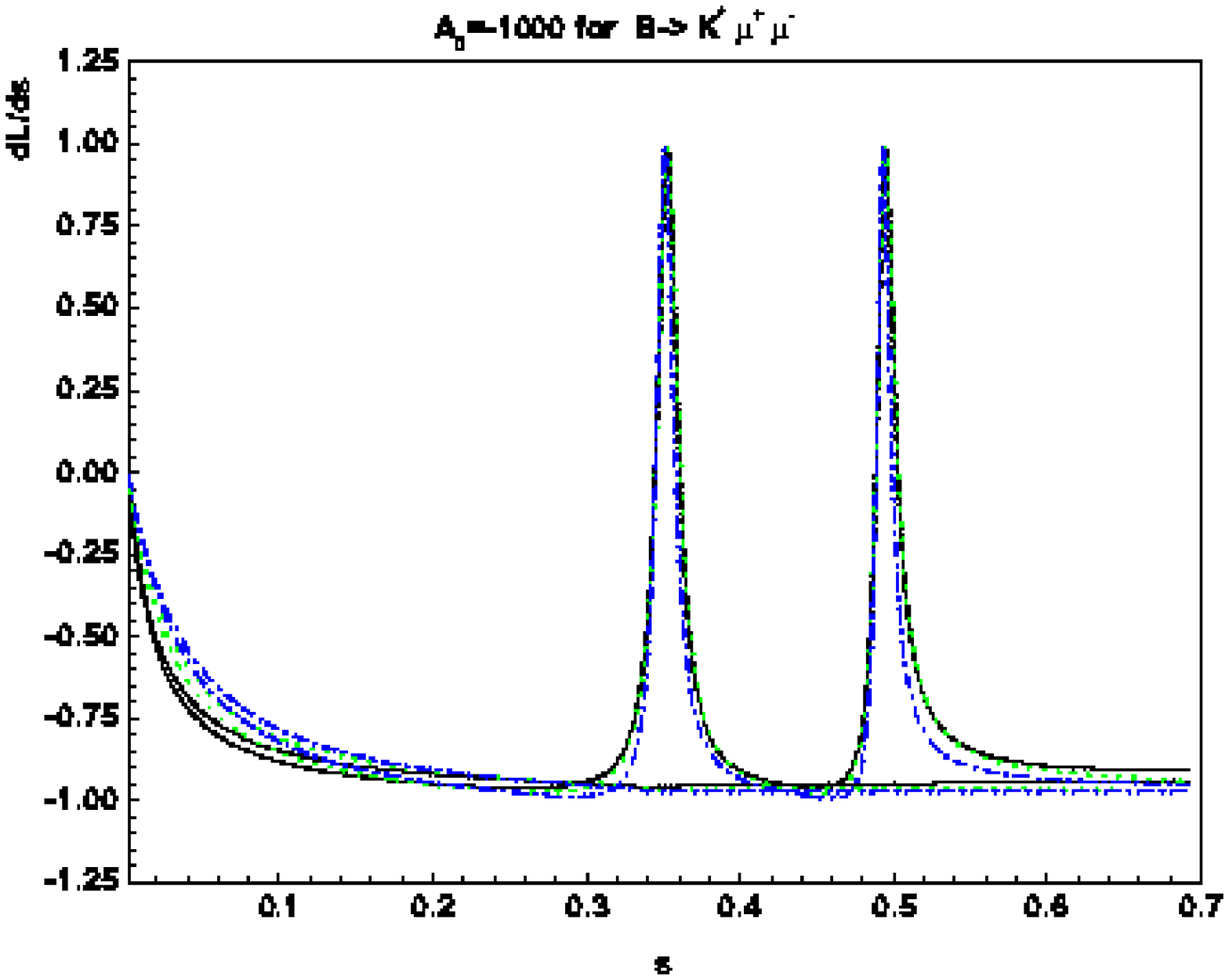}}
{\includegraphics[width=8.9cm] {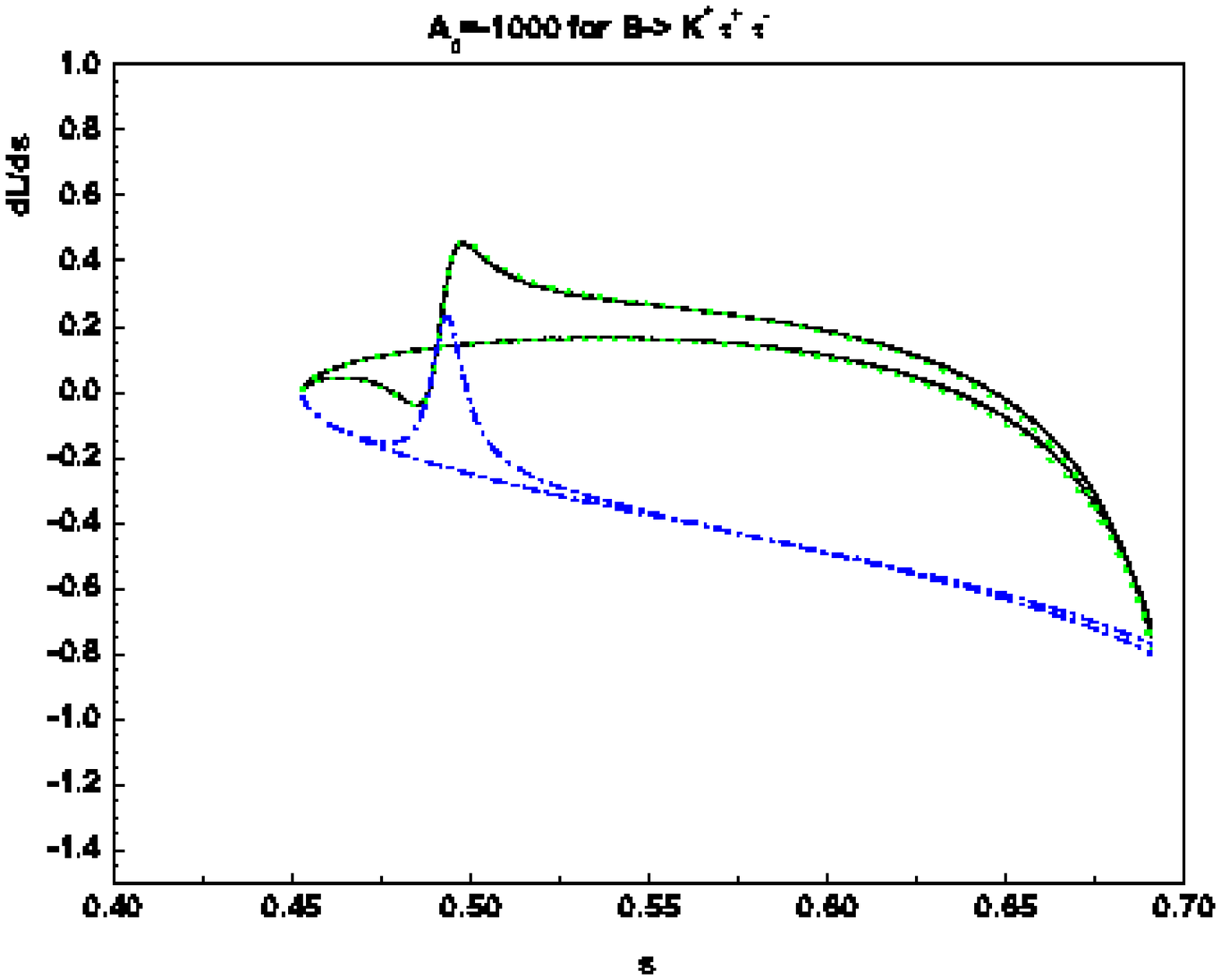}} \caption{
\label{bkL} The $dL/d\hat{s}$ of the process $B \to K^{(*)} l^+
l^-$ for $A_0= -1000$. The line conventions are the same as those
in Fig. \ref{bkims}.}
\end{figure}

\begin{figure}
{\includegraphics[width=8.7cm] {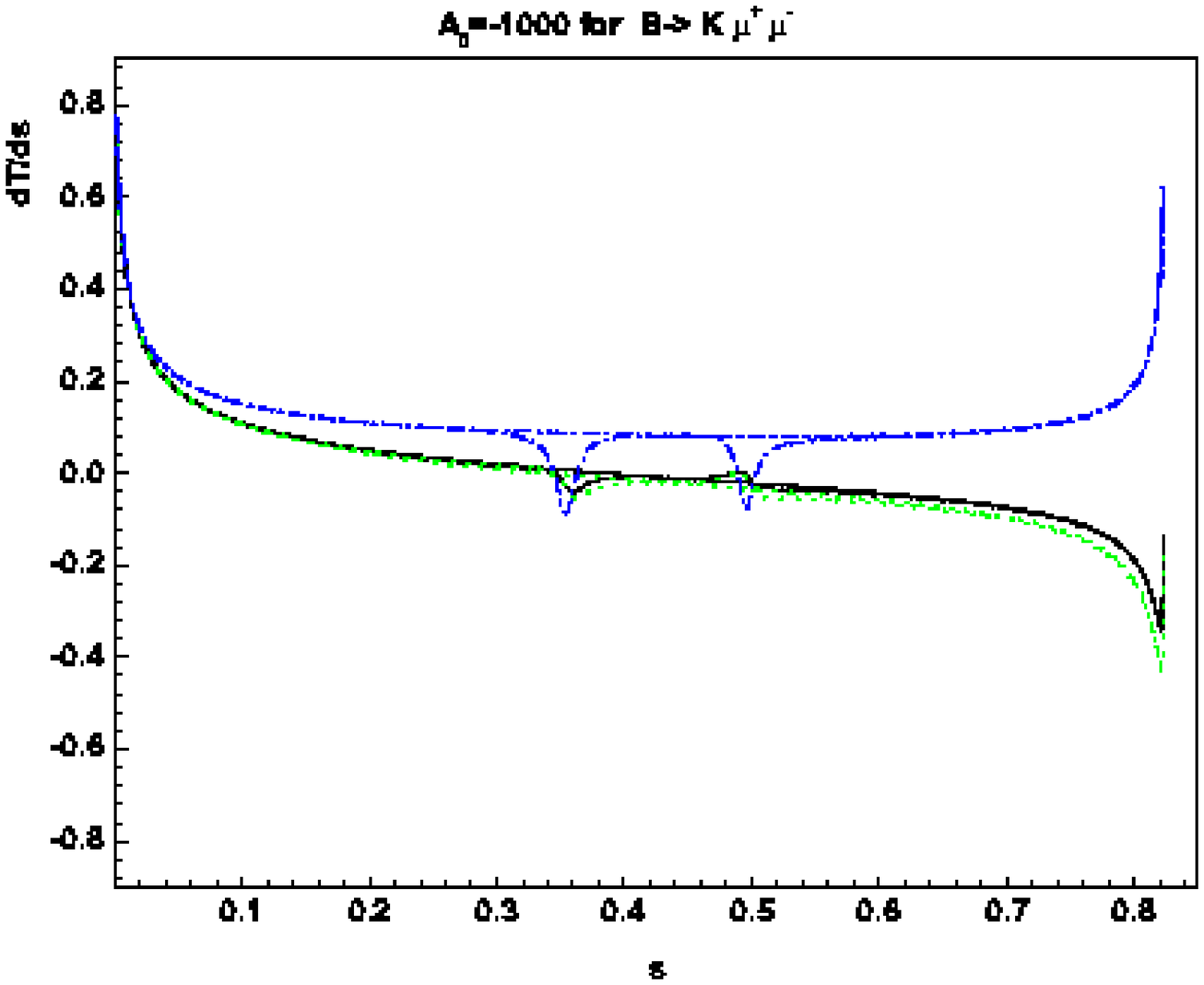}}
{\includegraphics[width=8.9cm] {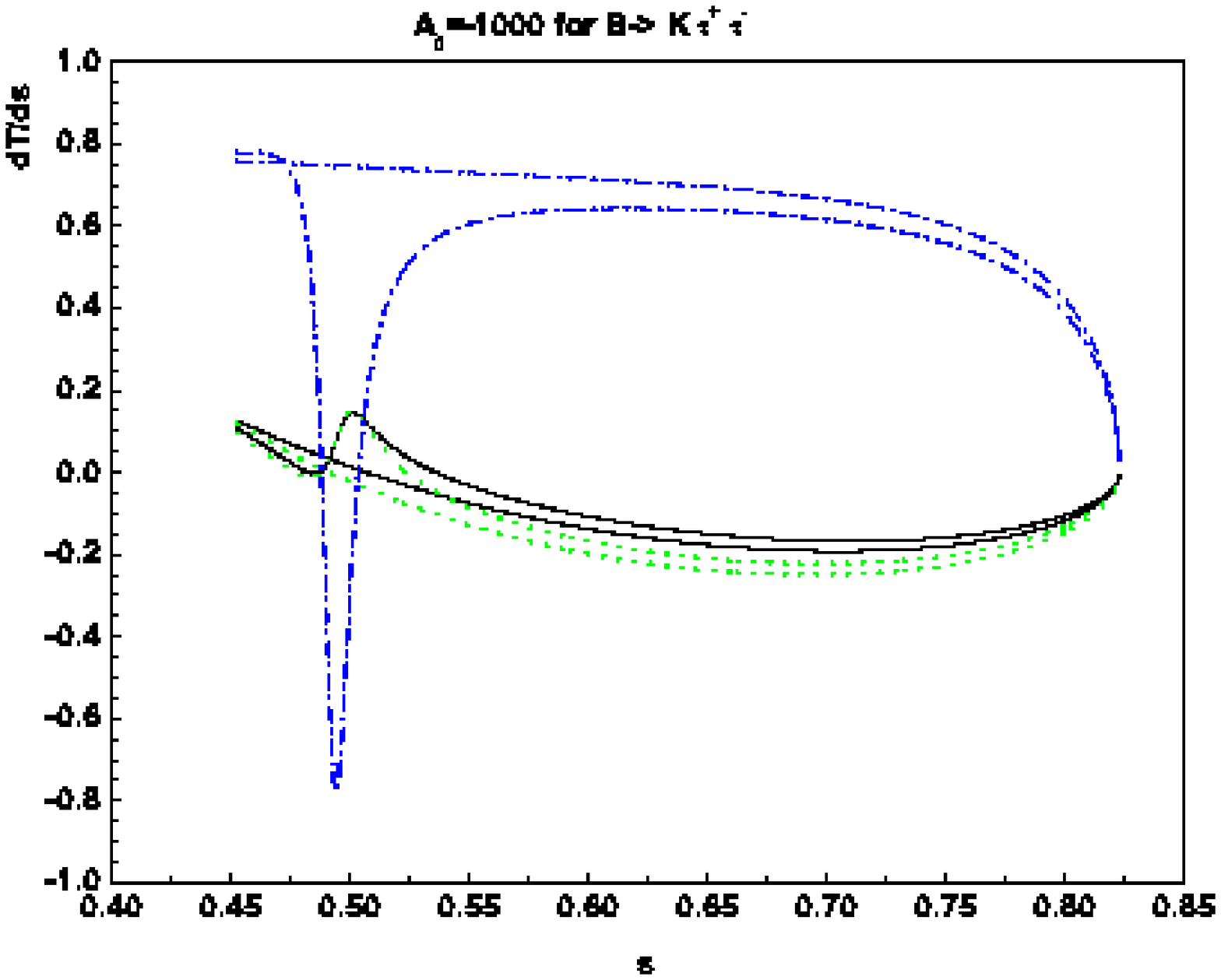}}
{\includegraphics[width=8.9cm] {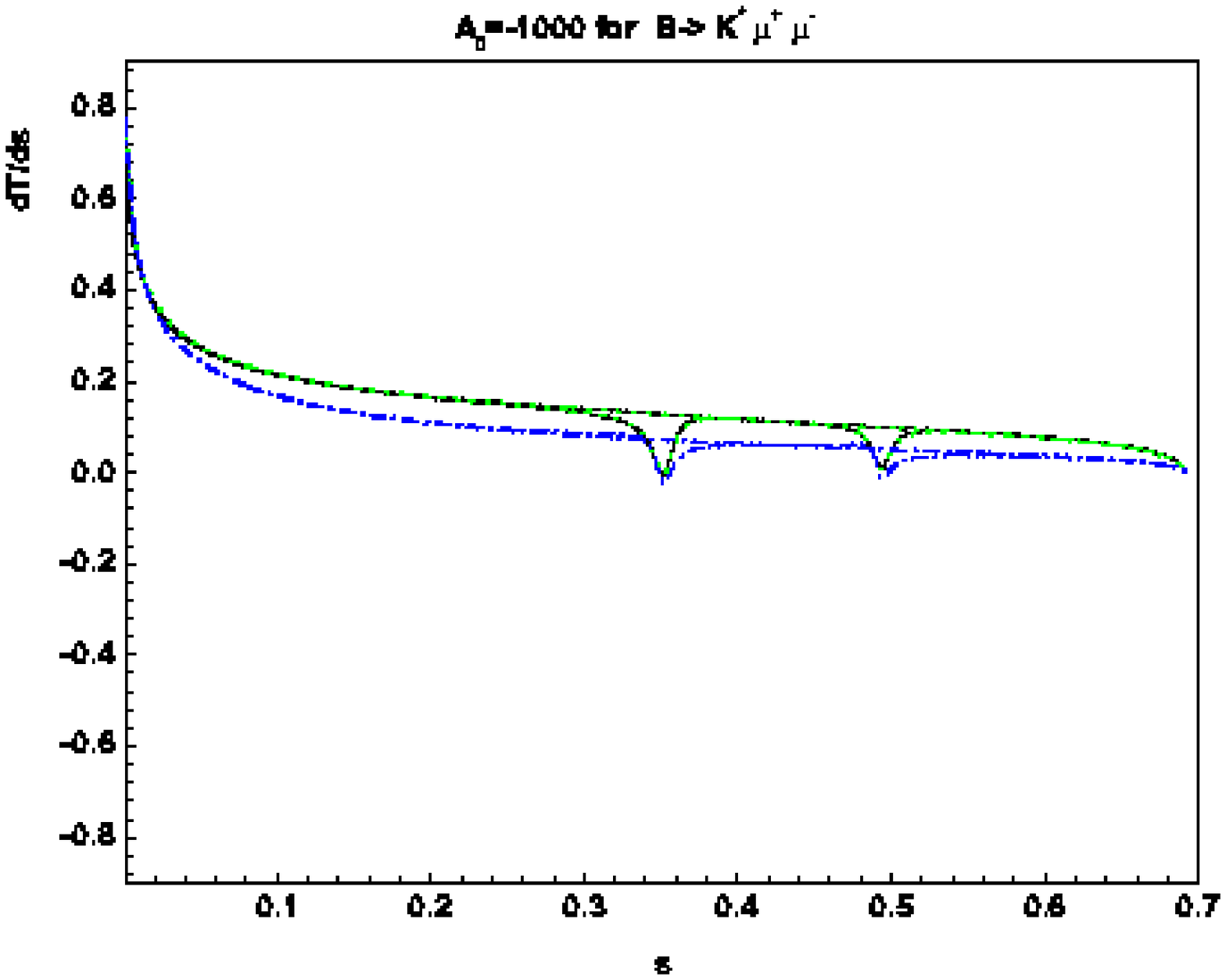}}
{\includegraphics[width=8.9cm] {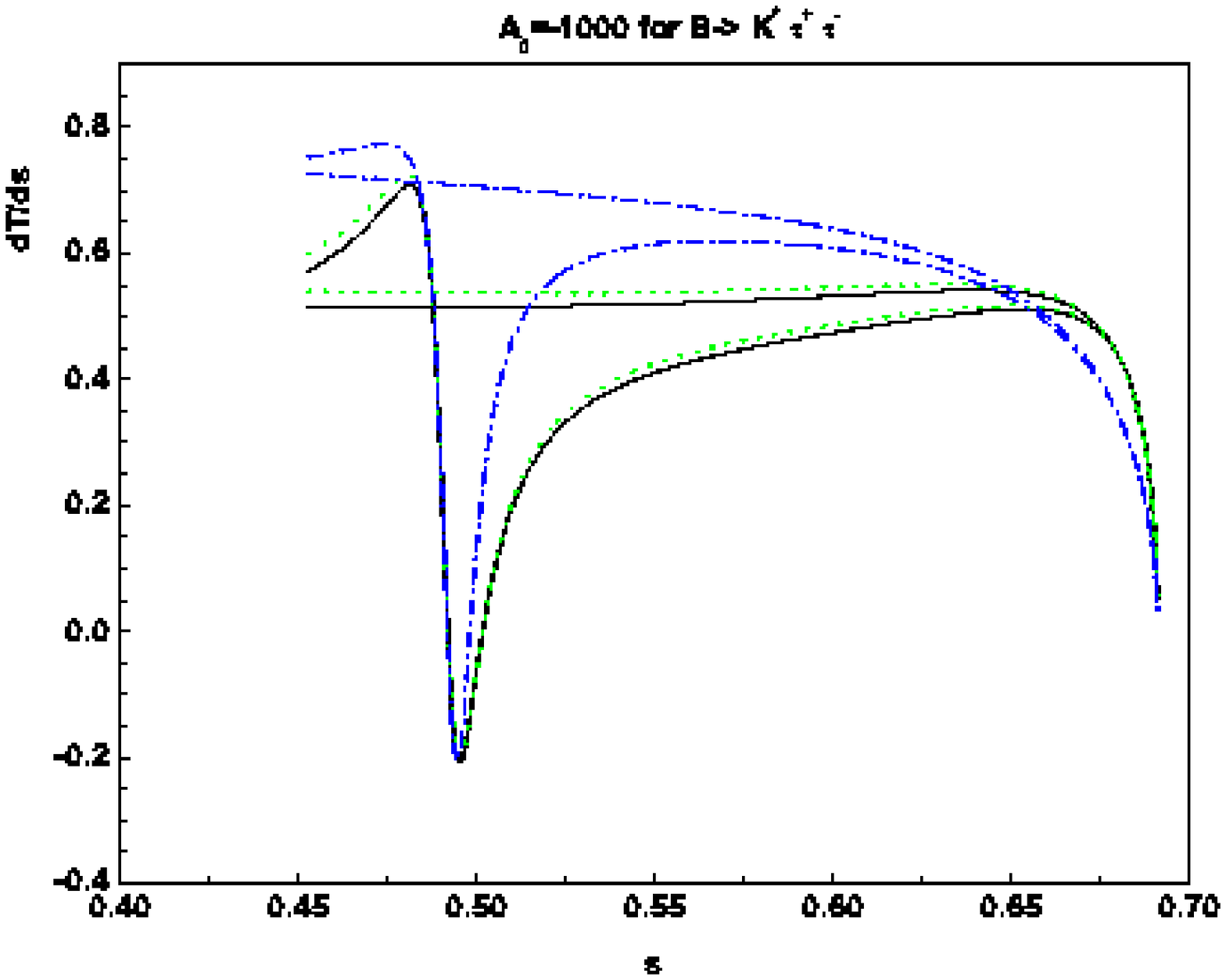}} \caption{
\label{bkT} The $dT/d\hat{s}$ of the process $B \to K^{(*)} l^+
l^-$ for $A_0= -1000$. The line conventions are the same as those
in Fig. \ref{bkims}.}
\end{figure}

\begin{figure}
{\includegraphics[width=8.7cm] {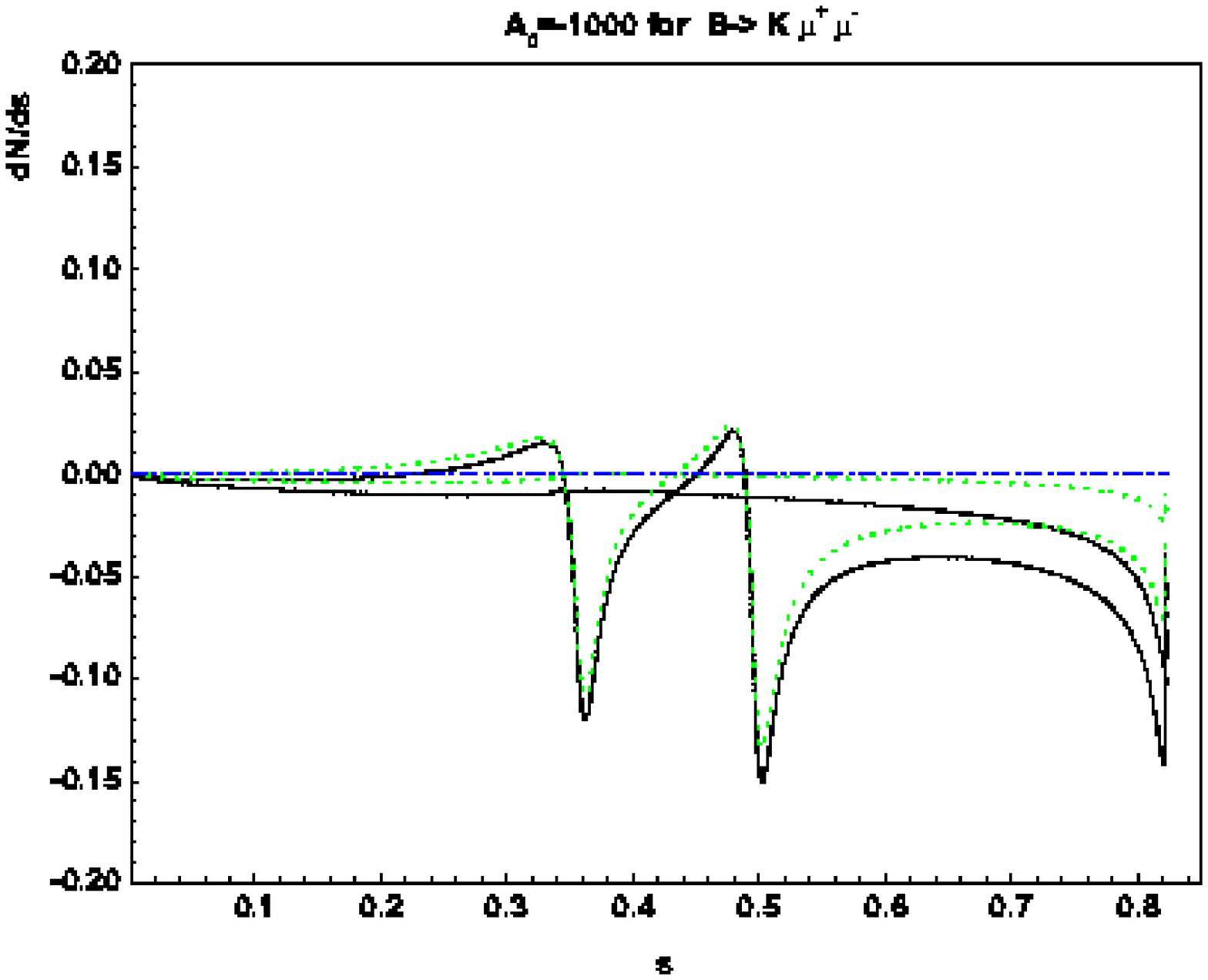}}
{\includegraphics[width=8.9cm] {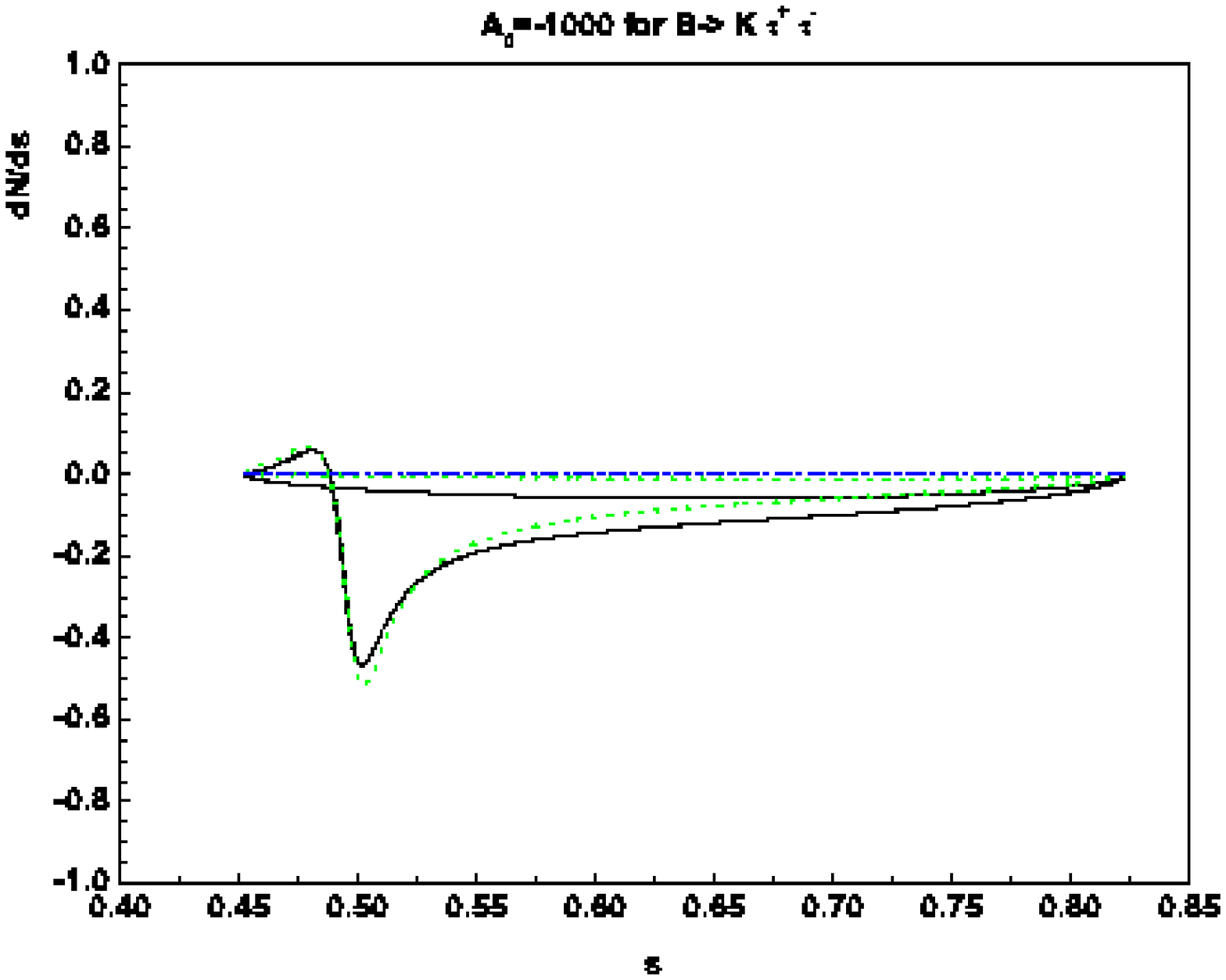}}
{\includegraphics[width=8.9cm] {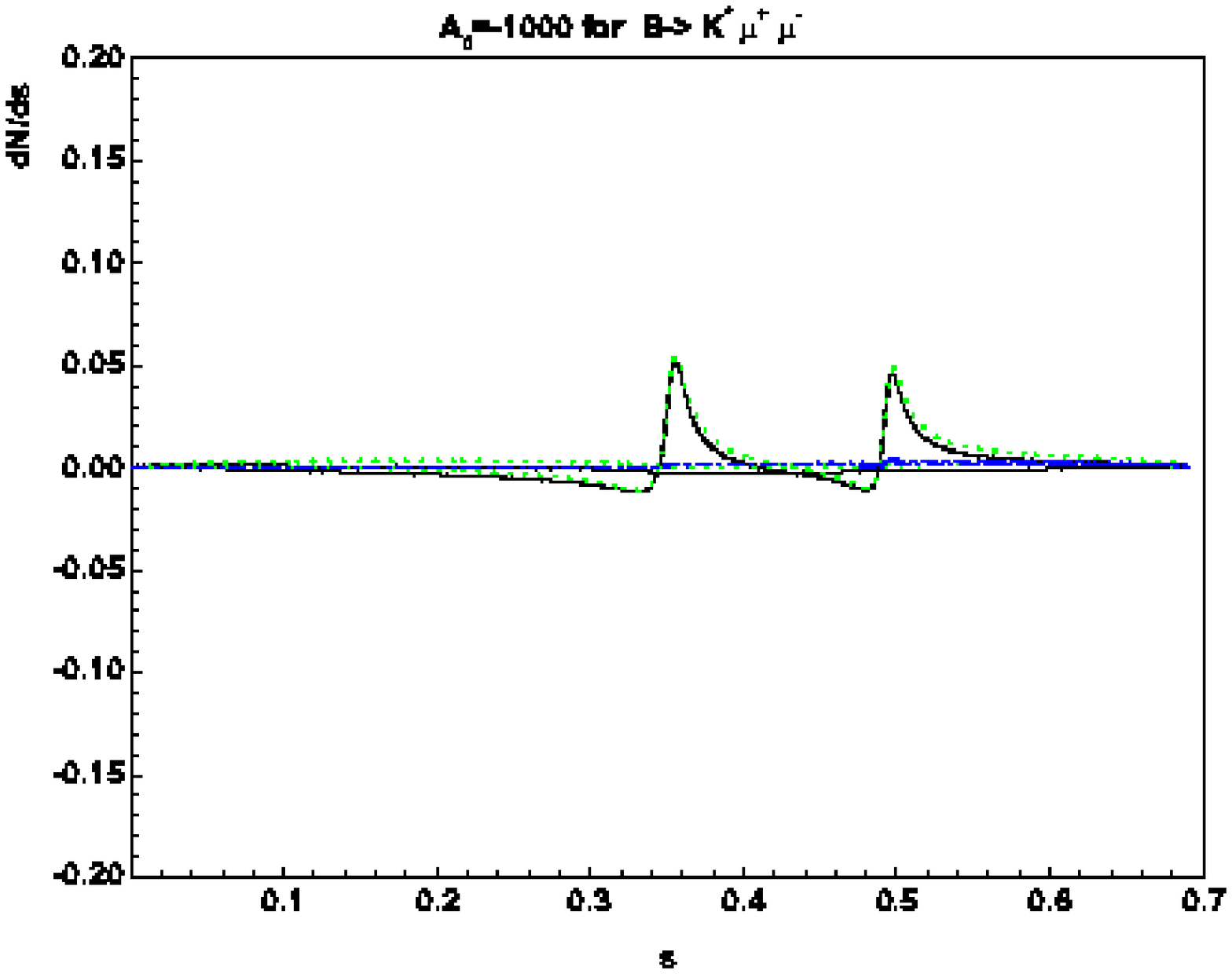}}
{\includegraphics[width=8.9cm] {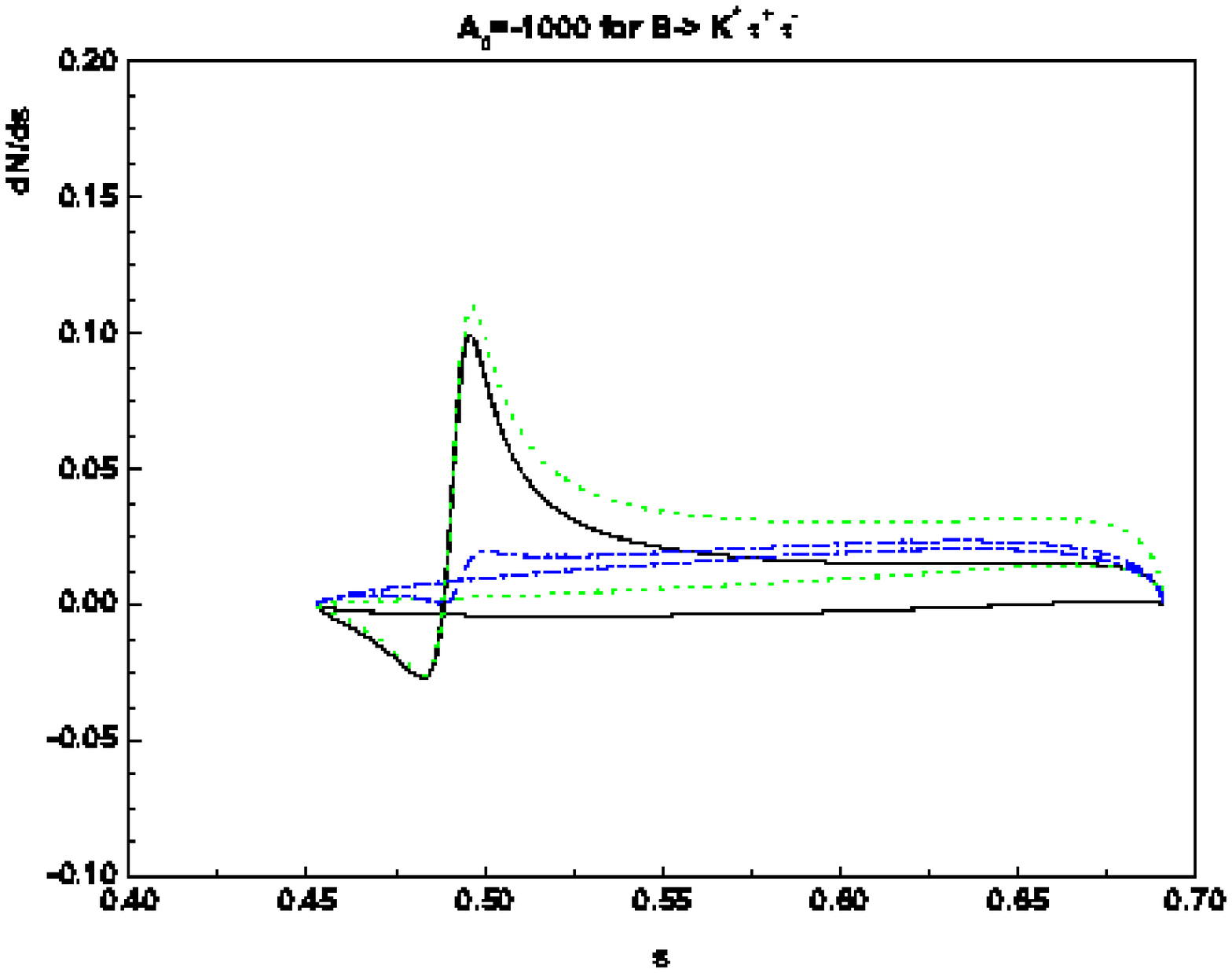}} \caption{
\label{bkN} The $dN/d\hat{s}$ of the process $B \to K^{(*)} l^+
l^-$ for $A_0= -1000$. The line conventions are the same as those
in Fig. \ref{bkims}.}
\end{figure}

The figures for the dependence of observable on $\hat{s}$ with and
without long-distance contributions are presented in {\bf
Fig}.(\ref {bkims}-\ref{bkN}) in the case of $A_0=-1000$, where
the solid lines denote the all contributions ($W^{\pm},H^{\pm}$,
chargino, gluino, neutrilino propagated in the loop) including the
NHB contributions; the dot lines present the SM contribution plus
only the NHB contributions, and the dot-dashed lines are for the
SM contribution. We also calculated the dependence of observable
on $\hat{s}$ in the case of $A_0=0$ and give the results for
$A_0=0$ when the two cases have a sizable difference.

    The IMS of the process $B \to K^{(*)} l^+ l^-$ is
given in {\bf Fig}.\ref {bkims}, where the left two figures are
for $B \to K^{(*)} \mu^+ \mu^-$ and the right for $B \to K^{(*)}
\tau^+ \tau^-$. For the case of $B \to K \mu^+ \mu^-$, we can see
that, at the low $\hat{s}$ region, there are some enhancement from
NHB contributions. Compared to the decay $B \to K \mu^+ \mu^-$,
the IMS of $B \to K^{*} \mu^+ \mu^-$ deviates from the SM
prediction sizably in the whole region of $\hat{s}$. Nevertheless,
the SUSY effects are small compared with the SM for the decay $B
\to K^{(*)} \tau^+ \tau^-$.

The {\bf Fig}. \ref{bkfba} is for the FBA ($dA/d\hat{s}$) of the
decay $B \to K^{(*)} l^+ l^-$, where the left two figures are for
$l=\mu$ and the right for $l=\tau$, like that in Fig.\ref{bkims}.
As it is known, the FBA ($dA/d\hat{s}$) of $B \to K l^+ l^-$ in
the SM is zero. Since FBA arises in the SUSY models only when NHB
effects are taken into account, it provides a good probe to test
these effects. Our numerical results show that the average of FBA
in $B \to K \mu^+ \mu^-$ can reach only 0.001 which is too small
to be observed. The average of FBA in $B \to K \tau^+ \tau^-$ can
reach -0.1 and 0.05 for the case of $A_0=0$ and $A_0=-1000$,
respectively. (The reason why FBA in the two cases has the
opposite sign is that the sign of function $S_1$ in these two
cases is opposite.) So, the $10^{10}-10^{11}$ $B_d\;\bar{B}_d$
pairs per year, which is in the designed range in the future super
B factors with B hadrons $10^{10}-10^{12}$ per year~\cite{superb},
is needed in order to observe the FBA with good accuracy.
Our results show that the SUSY effects show up at the low
$\hat{s}$ region for the FBA of $B\to K^{*} \mu^+ \mu^-$ and the
deviation from SM is 0.05 or so.   
It is worth to note that there is a sizable change of the position
of the zero-point of the FBA in $B\to K^* \mu^+\mu^-$ in the SUSY
SO(10) model, as it can be seen in Fig.5, which could be tested in
the future experiments with high precision. For FBA in $B \to
K^{*} \tau^+ \tau^-$, the deviation from the SM is about several
percent. The average of FBA of $B \to K^{*} \tau^+ \tau^-$ can
reach 0.3 in the case of $A_0=0$. To observe the FBA in $B \to
K^{*} \tau^+ \tau^-$ decay at 1$\sigma$ level, the required number
of events is $1.1 \times 10^8$. The number of B$\bar{B}$ pairs
that is expected to be produced at B factories is about $N \simeq
5 \times 10^8$. Therefore the FBA in $B \to K^{*} \tau^+ \tau^-$
could be observed at B factories. Hence, with the enhancement of
experimental precision and statistics, the measurements of FBA
would provide more data and effectively pin to the NP effects.

  Now, we turn to discuss the lepton polarization. We present the
longitudinal, transverse and normal polarization in the {\bf
Fig}.(\ref{bkL},\ref{bkT},\ref{bkN}) for $B \to K^{(*)} l^+ l^-$
decay. As it can be seen in {\bf Fig}. \ref{bkL} and \ref{bkT},
the $dL/d\hat{s}$ of $B \to K^{(*)} \mu^+ \mu^-$ is not sensitive
to the NHB effects, while for $dT/d\hat{s}$ of $B \to K (K^{*})
\mu^+ \mu^-$, the deviation from SM can reach 0.1 (0.05). As it is
expected, the contribution from the $\tau^+ \tau^-$ channel is
much larger than that from the $\mu^+ \mu^-$ one. For $B \to
K^{(*)} \tau^+ \tau^-$, the NHB contributions are manifest and
dominant, and both $dL/d\hat{s}$ and $dT/d\hat{s}$ are
significantly different from SM. And the $dL/d\hat{s}$ of $B \to K
\tau^+ \tau^-$ can even reach 0.6. Thus, the NHB effects are
sensitive to $B \to K^{(*)} \tau^+ \tau^-$ and will be observable
at B factories.

The $dN/d\hat{s}$ of $B \to K^{(*)} l^+ l^-$ decay are given in
{\bf Fig}. \ref{bkN}.  The average of $dN/d\hat{s}$ can reach
several percent for $B \to K \mu^+ \mu^-$ which could be observed
in the future super B factories, while it is the order of
$10^{-3}$ for $B \to K^{*} \mu^+ \mu^-$ which can not be observed
even in designed super B factories. The average of $dN/d\hat{s}$
in $B\to K \tau^+\tau^-$ is 0.05 or so. For $B \to K^{*} \tau^+
\tau^-$, the deviation from SM is a few percent. As noted above,
the Wilson coefficient $C_{10}$ is real and $C_{10}^\prime,
C_{Q_i}^{(\prime)}=0$ (precisely speaking, they are negligibly
small) in the SM so that $dN/d\hat{s}=0$ in $B \to K \mu^+ \mu^-$
in the SM. It is still true in the minimal super gravity model
(mSUGRA) and SUSY models with real universal boundary conditions
at the high scale~\cite{0004262}. In the SUSY SO(10) model we
considered, the complex flavor non-diagonal down-type squark mass
matrix element of 2nd and 3rd generations of order one at the GUT
scale induces the complex couplings which lead to the complex
Wilson coefficients and consequently the non zero normal
polarization of $B \to K \mu^+ \mu^-$. Therefore, the measurements
of the CP violating (as usual, the CPT invariance is assumed in
the paper) normal polarization in $B\to K l^+l^-$ could
discriminate the SUSY SO(10) model (and other SUSY models with the
flavor non diagonal complex couplings) from the SM and mSUGRA.

   In summary, we have carried out a study of SUSY effects, in
particular, the neutral Higgs bosons contributions to the IMS, FBA
and polarization, in the exclusive decay $B \to K^{(*)} l^+ l^-$
($l=\mu, \tau$) in the SUSY SO(10) model, taking account of the
constraints from existing experimental data such as $b \to s
\gamma$, $\Delta M_s$, $Br (B\to K^{(*)}\mu^+\mu^-)$, $\tau \to
\mu \gamma $ as well as the upper bound of $Br (B_s \to \mu^+
\mu^-)$. Our main findings can be summarized as follows:
\begin{itemize}
\item The IMS of the process $B \to K^{(*)} \mu^+ \mu^-$ can
sizably deviate from the SM.  \item The FBA comes only from NHB
contributions in $B \to K l^+ l^-$ and its average for $l=\mu$ is
nonzero but too small to be ovserved. However for $B \to K \tau^+
\tau^-$, it is the order of 10$\%$, which should be within the
luminosity reach of coming B factories. The SUSY effects show up
at the low $\hat{s}$ region for the FBA of $B\to K^{*} \mu^+
\mu^-$ and the deviation from SM is 0.05 or so. Moreover, there is
a sizable change of the position of the zero-point of the FBA in
$B\to K^* \mu^+\mu^-$, which can be used to discriminate the model
from the SM. \item The average of $dN/d\hat{s}$ can reach several
percent for $B \to K \mu^+ \mu^-$ and it is 0.05 or so for $B\to K
\tau^+\tau^-$, which could be measured in the future super B
factories and provide a useful information to probe new physics
and discriminate different models. \item The longitudinal
polarization, $dL/d\hat{s}$, of $B \to K^{(*)} \mu^+ \mu^-$ is not
sensitive to the NHB effects. However, for the transverse
polarization, $dT/d\hat{s}$, of $B \to K (K^{*}) \mu^+ \mu^-$, the
deviation from SM can reach 0.1 (0.05) which could be seen in B
factories. For $B \to K^{(*)} \tau^+ \tau^-$, the NHB
contributions are manifest and dominant, and both $dL/d\hat{s}$
and $dT/d\hat{s}$ are significantly different from SM. And the
$dL/d\hat{s}$ of $B \to K \tau^+ \tau^-$ can even reach 0.6, which
can be measured in B factories.
\end{itemize}

Therefore, the experimental investigation of observable, in
particular, FBA and the polarization components, in the $B \to
K^{(*)} l^+ l^-$ decays in the present B factories and future
super B factories can be used to search for SUSY effects, in
particular, NHB effects, in SUSY grand unification models.

\section*{Acknowledgement}

 One of the authors (W.-J. Li) would like to thank Dr. X.-H. Wu for
discussions during the work. The work was supported in part by the
National Nature Science Foundation of China.


\section*{References}
\begin{thebibliography}{99}
\bibitem{ichep}
A.Ishikawa, (Belle Collaboration), Lecture given at ICHEP 2004.
\bibitem{Aubert}A. Ishikawa {\it et. al}, BELLE Collaboration,
Phys. Rev. Lett. {\bf 91}, 261601 (2003);
 B. Aubert {\it et al.}, (BABAR Collaboration), Phys. Rev. Lett.
{\bf 91}, 221802 (2003).
\bibitem{quarkmodel}
P. Ball and R. Zwicky, hep-ph/0406232; M.A. Ivanov and V.E.
Lyubovitskij, Lectures given at International School on Heavy
Quark Physics, Dubna, Russia, 27 May - 5 Jun 2002; W. Jaus and D.
Wyler, Phys. Rev. {\bf D 41} (1990) 3405; P. Colangelo {\it et
al.}, Phys. Lett. {\bf B317} (1993) 183; Ceng {\it et al.}, Phys.
Rev. {\bf D 54} (1996) 3656; Amand Faessler {\it et al.}, EPJ C,
4, {\bf C18}, 1每33 (2002).
\bibitem{svzqcd}
P. Colangelo {\it et al.}, Phys. Rev. {\bf D 53} (1996) 3672
[arXiv:hep-ph/9510403]; Erratum, ibid, {\bf D 57} (1998) 3186.
\bibitem{lcsr}
I.I. Balitsky, V.M. Braun and A.V. Kolesnichenko, Nucl. Phys. {\bf
B312} (1989) 509; V.L. Chernyak and I.R. Zhitnitsky, Nucl. Phys.
{\bf B345} (1990) 137; V.M. Braun, Preprint NORDITA每98每1每P
[arXiv:hep-ph/9801222]; A. Khodjamirian and R. Ruckl,
[arXiv:hep-ph/9801443]; P. Ball and V.M. Braun, Phys. Rev. {\bf D
58} (1998) 094016 [arXiv:hep-ph/9805422]; Nucl. Phys. {\bf B543}
(1999) 201 [arXiv:hep-ph/9810475]; P. Ball {\it et al.}, Nucl.
Phys. {\bf B529} (1998) 323 [arXiv:hep-ph/9802299];
 P. Ball, JHEP {\bf 09} (1998) 005 [arXiv:hep-ph/9802394];
  JHEP {\bf 01} (1999) 010 [arXiv:hep-ph/9812375].
\bibitem{rcqm}
C.Q. Geng and C.P. Kao, Phys. Rev. {\bf D 54} (1996) 5636
[arXiv:hep-ph/9608466]; Phys. Rev. {\bf D 57} (1998) 4479.
\bibitem{modindep}
S. Fukaea, C.S. Kimb, T. Morozumi, and T. Yoshikawac,
Phys. Rev. {\bf D 59} (1999) 074013 [arXiv:hep-ph/9807254]; T.M.
Aliev,C. S. Kim and Y. G. Kim, Phys. Rev. {\bf D 62}, 014026
(2000) [arXiv:hep-ph/9910501]; T. M. Aliev,et.al, Nucl. Phys. {\bf
B607} (2001) 305-325 [arXiv:hep-ph/0009133]; T.M. Aliev, M.K.
Cakmak, A. $\ddot{O}$zpineci, M. Savci, Phys. Rev. {\bf D 64}
(2001) 055007 [arXiv:hep-ph/0103039].
\bibitem{dan}
Benjamin Grinstein, Dan
Pirjol,CTP-MIT-3490,[arXiv:hep-ph/0404250].
\bibitem{0003188}
T.M. Aliev, M. Savci, Phys. Lett. {\bf B481} (2000) 275-286
[arXiv:hep-ph/0003188].
\bibitem{0112149}
L.T. Handoko, C.S. Kim, T. Yoshikawa, Phys. Rev. {\bf D 65} (2002)
077506 [arXiv:hep-ph/0112149].
\bibitem{0112300}
A. Ali, E. Lunghi, C. Greub, G. Hiller, DESY 01-217, BUTP-01-21,
SLAC-PUB-9076 [arXiv:hep-ph/0112300].
\bibitem{9705222}
T.M. Aliev, M. Savci, A. Ozpineci, H. Koru,J.Phys. G24 (1998)
49-65 [arXiv:hep-ph/9705222].
\bibitem{0004262}Qi-Shu Yan, Chao-Shang Huang, Liao Wei,
Shou-Hua Zhu, Phys. Rev. {\bf D 62} (2000) 094023
[arXiv:hep-ph/0004262].
\bibitem{9910221}
A. Ali, P. Ball, L.T. Handoko, G. Hiller, Phys. Rev. {\bf D 61}
(2000) 074024 [arXiv:hep-ph/9910221].
\bibitem{0008210}
 F. Kr$\ddot{u}$ger, E. Lunghi, Phys. Rev. {\bf D 63} (2001) 014013 [arXiv:hep-ph/0008210].
\bibitem{0009149-2}
C.S. Huang, Nucl.Phys. Proc. Suppl. {\bf 93} (2001) 73-78
[arXiv:hep-ph/0009149]; C. Bobeth, T. Ewerth, F. Kruger, J. Urban,
Phys. Rev. {\bf D 64} (2001) 074014 [arXiv:hep-ph/0104284].
\bibitem{0204219cpp}
Guray Erkol, Gursevil Turan, Nucl. Phys. {\bf B635} (2002) 286-308
[arXiv:hep-ph/0204219].
\bibitem{0307276FBA}
S. R. Choudhury {\it et al.}, Phys. Rev. {\bf D 69} (2004) 054018
[arXiv:hep-ph/0307276].
\bibitem{0304084}
S. Rai Choudhury {\it et al.}, Phys. Rev. {\bf D 68}, 054016
(2003) [arXiv:hep-ph/0304084]
\bibitem{0209228}
 Wafia Bensalam, David London, Nita Sinha, Rahul Sinha,
Phys. Rev. {\bf D 67} (2003) 034007 [arXiv:hep-ph/0209228]
\bibitem{double} T.M.
Aliev, V. Bashiry, M. Savci, Eur. Phys. J. {\bf C35} (2004)
197-206 [arXiv:hep-ph/0311294]; JHEP {\bf 0405} (2004) 037
[arXiv:hep-ph/0403282]; A.S. Cornell, Naveen Gaur,
[arXiv:hep-ph/0408164].
\bibitem{neu}
V. Barger {\it et al.}, Phys. Rev. Lett. {\bf 82} (1999) 2640
[arXiv:astro-ph/9810121]; E. lisi {\it et al.}, Phys. Rev. Lett.
{\bf 85} (2000) 1166 [arXiv:hep-ph/0002053]; John N. Bahcall {\it
et al.}, Phys. Rev. Lett. {\bf 90}, 131301 (2003)
[arXiv:astro-ph/0212331]; John N. Bahcall and M. H. Pinsonneault,
Phys. Rev. Lett. {\bf 92}, 121301 (2004) [arXiv:astro-ph/0402114];
Sandhya Choubey and Probir Roy, Phys. Rev. Lett. {\bf 93}, 021803
(2004) [arXiv:hep-ph/0310316].
\bibitem{asy}
K.S. Babu and S.M. Barr, Phys. Lett. {\bf B381} (1996) 202
[arXiv:hep-ph/9511446]; C.H. Albright, K.S. Babu, and S.M. Barr,
Phys. Rev. Lett. {\bf 81} (1998) 1167 [arXiv:hep-ph/9802314]; J.
Sato and T. Yanagida, Phys. Lett. {\bf B430} (1998) 127
[arXiv:hep-ph/9710516]; N. Irges, S. Lavignac, and P. Ramond,
Phys. Rev. {\bf D 58} (1998) 035003 [arXiv:hep-ph/9802334].
\bibitem{cmc1}
 Mu-Chun Chen, K.T. Mahanthappa,Phys.Rev.{\bf D62} 113007 (2000),
[arXiv:hep-ph/0005292].
\bibitem{0205111}
D. Chang, A. Masiero and H. Murayama, Phys. Rev. {\bf D67} (2003)
075013 [arXiv:hep-ph/0205111].
\bibitem{bsv}
B. Bajc, G. Senjanovi$\acute{c}$ and F. Vissani,
[arXiv:hep-ph/0210207]; H.S. Goh, R.N. Mohapatra and S.-P. Ng,
[arXiv:hep-ph/0303055].
\bibitem{bi}
X-J. Bi, Y-B. Dai and X-Y Qi, Phys. Rev. {\bf D63} 096008 (2001);
X-J. Bi and Y-B. Dai, Phys. Rev. {\bf D66} 076006 (2002).
\bibitem{cmc2}
Mu-Chun Chen, K.T. Mahanthappa, Phys.Rev.{\bf D65} 053010 (2002),
[arXiv:hep-ph/0106093]; Mu-Chun Chen, K.T.
Mahanthappa,Phys.Rev.{\bf D68} 017301
(2003),[arXiv:hep-ph/0212375];Mu-Chun Chen and K.T. Mahanthappa,
Int.J.Mod.Phys. {\bf A18}:5819-5888,2003
[arXiv:hep-ph/0305088];Mu-Chun Chen, K.T.
Mahanthappa,[arXiv:hep-ph/0409096].
\bibitem{0312145}
Sebastian Jager, Ulrich Nierste,FERMILAB-Conf-03/394-T,
[arXiv:hep-ph/0312145].
\bibitem{0407263}
Y-B. Dai, C-S. Huang, W-J. Li, X-H. Wu, [arXiv:hep-ph/0407263].
\bibitem{goto}
T. Goto {\it et al.}, Phys. Rev. {\bf D 55} (1997) 4273; T. Goto,
Y. Okada and Y. Shimizu, Phys. Rev. {\bf D 58} (1998) 094006; S.
Bertolini, F. Borzynatu, A. Masiero and G. Ridolfi, Nucl. Phys.
{\bf B353} (91) 591.
\bibitem{dyb}
Y-B. Dai, C-S. Huang, H-W. Huang, Phys. Lett. {\bf
B390}:257-262,1997; Erratum-ibid. {\bf B513}:429-430, 2001
[arXiv:hep-ph/9607389].
\bibitem{wuxh}
Chao-Shang Huang, Xiao-Hong Wu, Nucl. Phys. {\bf B657} (2003)
304-332 [arXiv:hep-ph/0212220]; Jian-Feng Cheng, Chao-Shang Huang,
Xiao-Hong Wu, hep-ph/0404055, to appear in Nucl. Phys. B.
\bibitem{buras}
A.J. Buras, M. Muenz, Phys. Rev. {\bf D 52} (1995) 186-195
[arXiv:hep-ph/9501281].
\bibitem{Bs2mu}
D. Acosta {\it et al.}, (CDF Collaboration),
[arXiv:hep-ex/0403032].
\bibitem{bsmu}C.-S. Huang, W. Liao, Q.-S. Yan, S.-H. Zhu, Phys.
Rev. {\bf D63} (2001) 114021; ibid. {\bf 64} (2001) 059902(E)
[arXiv:hep-ph/0006250]; For review papers, see, e.g., C.-S. Huang,
Nucl. Phys. Proc. Suppl. {\bf 115}, 89 (2003); A. Dedes, Mod.
Phys. Lett. {\bf A18}, 2627 (2003) [arXiv:hep-ph/0309233].
\bibitem{d0}The $D_0$ collaboration, $D_0$ Coneference Note 4514.
\bibitem{msd}
A. Stocchi, Nucl. Phys. Proc. Suppl. {\bf 117}(2003) 145
[arXiv:hep-ph/0211245].
\bibitem{bsg}
A. Kagan, [arXiv:hep-ph/9806266]; T.E. Coan {\it et al.}, (CLEO
Collaboration), Phys. Rev. Lett. {\bf 80}, 1150 (1998)
[arXiv:hep-ex/9710028].
\bibitem{pdg}K. Hagiwara et al., Phys. Rev. {\bf D66}, 010001
(2002).
\bibitem{cmsvv}
M. Ciuchini {\it et al.}, Phys. Rev. Lett. {\bf 92} (2004) 071801.
\bibitem{taumugamma}K.Abe et. al., Belle collaboration,
Phys. Rev. Lett. {\bf 92} (2004) 171802 [arXiv:hep-ex/0310029].
\bibitem{superb}T.E. Browder and A. Soni, hep-ph/0410192, and
references therein.
\bibitem{nanop}J.L. Lopez, D.V. Nanopoulos, X. Wang and A. Zichichi,
Phys. Rev. {\bf D51}, 147 (1995) [arXiv:hep-ph/9406427].
\bibitem{hy}
C.-S. Huang and Q.-S. Yan, Phys. Lett. {\bf B442}(1998) 209
[arXiv:hep-ph/9803366]; C.-S. Huang, W. Liao, and Q.-S. Yan, Phys.
Rev. {\bf D59} (1999) 011701 [arXiv:hep-ph/9803460].
\bibitem{deltams}
 A. Stocchi, Nucl. Phys. Proc. Suppl. {\bf 117} (2003) 145 [arXiv:hep-ph/0211245].
\end{thebibliography}
\end{document}